\documentclass[amsmath,amssymb,prb,twocolumn,superscriptaddress,longbibliography]{revtex4-2}

\usepackage{graphicx} 
\usepackage[dvipsnames]{xcolor}
\usepackage{xcolor,colortbl}
\usepackage{bm}
\usepackage{hyperref}
\usepackage[final]{pdfpages}

\makeatletter
\patchcmd{\@outputpage@head}{\@ifx{\LS@rot\@undefined}{}{\LS@rot}}{}{}{}
\makeatother

\newcommand{\cfm}{Centro de F\'{\i}sica de Materiales (CFM) CSIC-UPV/EHU, E-20018, Donostia-San~Sebasti\'an, Spain}

\newcommand{\Figref}[1]{Fig.~\ref{#1}}
\newcommand{\Secref}[1]{Sec.~\ref{#1}}
\newcommand{\Eqref}[1]{Eq.~(\ref{#1})}

\newcommand{\ie}{\textit{i.e.}}
\newcommand{\eg}{\textit{e.g.}}
\newcommand{\SIfigNintTwentyFive}{Fig.~S5}
\newcommand{\SIfigWidths}{Fig.~S6}
\newcommand{\SIfigSVD}{Fig.~S7}
\newcommand{\SIfigRank}{Fig.~S8}
\newcommand{\SIfigU}{Fig.~S9}
\newcommand{\SIfigSpin}{Fig.~S10}
\newcommand{\SIfigSpinAFM}{Fig.~S11}
\newcommand{\SIfigSpinFM}{Figs.~S12-S15}
\newcommand{\SIfigStrainDFT}{Figs.~S16-S18}
\newcommand{\SIsecKREP}{Sec.~S1}
\newcommand{\SIsecNint}{Sec.~S2}

\newcommand{\SIsecEffective}{Sec.~S4}

\newcommand{\SIsecStrainDFT}{Sec.~S10}
\newcommand{\SIsecAnalyticalEffective}{Sec.~S11}

\begin{document}

\title{Predicting interface and spin states in armchair graphene nanoribbon junctions}

\author{Sofia Sanz}
\email{sofia.sanzwuhl@ehu.eus}
\affiliation{\cfm}

\author{Daniel S\'anchez-Portal}
\email{daniel.sanchez@ehu.eus}
\affiliation{\cfm}

\date{\today}

\begin{abstract}
	We present a theoretical analysis of interface states emerging at junctions between armchair graphene nanoribbons of varying widths. By exploring diverse width combinations and junction geometries, we demonstrate that predicting the precise number of interface states requires considerations beyond the topological classification alone; specifically, the width differences and bonding configuration at the interface play crucial roles.
    For junctions involving ribbons with small gaps, we further examine how an applied strain affects their topological properties and, consequently, the interface states formed.
    The spin states at these junctions are investigated using the mean-field Hubbard model, revealing how the magnetic behavior at the interface depends on the number of localized states present.
    These results are summarized in a series of ``rules of thumb" to predict the number of localized states and the magnetic moment at the junction.
    Our findings contribute to understanding and engineering localized states in graphene-based devices, providing guidelines for manipulating electronic and magnetic properties through structural design.
\end{abstract}

\maketitle

\section{Introduction}

Graphene nanoribbons (GNRs) have emerged as key building blocks for advanced nanoelectronic, spintronic, and optoelectronic devices, owing to their unique physical properties, such as highly tunable band structures and edge-dependent electronic states  \cite{Nakada1996, Wakabayashi1999, Rocha2005, Son2006b, Brey2006, Son2006a, Han2007, Areshkin2007, Yang2007, Dubois2009, Schwierz2010, Jaskolski2011, Bennett2013, Cummings2017, Zhang2023}.
In addition to their electronic properties, the advances in experimental techniques enable the fabrication of atomically precise GNRs with tailored widths  \cite{Cai2010, Ruffieux2016}. 
Besides their intrinsic properties, these laterally confined, semiconducting strips of graphene exhibit a fascinating topological nature, which has become a major area of focus in recent years \cite{Cao2017, Groening2018, Rizzo2018, Jiang2021, LopezSancho2021, Zdetsis2021, Li2021, Zdetsis2023}.
In the seminal paper by Cao et al., the authors showed for the first time that the topological phase of armchair GNRs (AGNRs) with spatial symmetry is dictated by their width, edge shape and terminating unit cell \cite{Cao2017}.
In this regard, GNRs can be characterized by a $Z_2$ invariant which takes a value of either 0 or 1 and classifies them as topologically trivial or non-trivial, respectively \cite{Fu2007, Cao2017}. 
Topologically non-trivial semiconducting AGNRs can host highly robust symmetry-protected in-gap states localized at the ribbon terminations, while maintaining an insulating bulk \cite{Cao2017}. 
%
%
Later, the $Z_2$ invariant was generalized into the chiral-phase index ($Z$), which relies on the chiral symmetry rather than time-reversal symmetry, providing a broader classification scheme for GNR topological phases \cite{Jiang2021}.
It is now well-established that the width, length, and termination geometry of AGNRs strongly influence their topological invariants ($Z_2$ and $Z$) and the emergence of localized in-gap 
end states (ESs)~\cite{Zdetsis2020, Lawrence2020, Zdetsis2021, LopezSancho2021,Zdetsis2023, GarciaFuente2023}.
For instance, wider AGNRs host multiple topological transitions, resulting in the appearence of multiple ESs \cite{LopezSancho2021, GarciaFuente2023, Zdetsis2023}, where the topological invariant $|Z|$ corresponds to the number of these states \cite{Jiang2021, LopezSancho2021}.

The interest on the topology of single GNRs, has naturally evolved to the novel quantum phases that emerge by coupling sections of GNRs of different topological character.
For instance, AGNR heterojunctions provide a platform to explore rich physical phenomena arising from the interplay of topology and geometry, since interfaces between topologically distinct AGNRs host half-filled, localized bound states within the energy gap \cite{PhysRevB.78.245402, Cai2014, Chen2015, Wang2017, Rizzo2018, Groening2018, Lee2018, Lin2018, Joost2019, Rizzo2020, Tamaki2020, Ostmeyer2024}.
In this sense, similarly to single GNRs, AGNR superlattices can be classified into topological or trivial junctions, where the key ingredient is the difference between the widths of the AGNRs \cite{Jiang2021, LopezSancho2021}.
However, while the topological classification provides valuable insight into the \emph{existence} of interface states, it does not fully determine their \emph{precise} number, which have received much less attention.
For instance, while it has been suggested that bonding at the interface plays a crucial role \cite{Lv2019} in the formation of interface states, prior studies typically focus on fixed bonding configurations and lack a systematic exploration of how variations in bonding affect the formation of localized states. 
Likewise, strain, is known to strongly influence key electronic properties such as magnetism and topology in GNRs \cite{Li2010a, Lawrence2020, Moles2025, SANTOS20201, Huang2024}.
Moreover, the magnetic properties associated with these localized states at the interface, essential for spintronic applications, have not been comprehensively addressed.

We bridge these gaps by performing a systematic theoretical study of AGNR heterojunctions of varying widths, alignments, and some geometrical deformations.
In contrast to previous works mainly emphasizing the topological classification, here we focus on the precise counting of interface states. We show that the number of interface states is set by the hybridization of the AGNR ESs through the specific bonding pattern, which defines a coupling matrix not captured by topology alone.
Using a combination of tight-binding (TB) and mean-field Hubbard (MFH) calculations, we explore the role of interface bonding, topology and width difference, 
while also investigating the effects of strain and Coulomb interactions.
The computed spin configurations at the junction are benchmarked against density functional theory (DFT) calculations,  which confirm the validity of our approach. From these results, we derive predictive ``rules of thumb" that relate interface structure to the number of interface states, providing useful design guidelines for AGNR-based circuits with tunable electronic and magnetic properties.

This paper is organized as follows: In \Secref{sec:methods}, we describe the Hamiltonian used to model the systems under investigation.
In \Secref{sec:end-states-AGNRs} we rationalize the emergence of ESs in AGNRs as a function of the ribbon width.
In \Secref{sec:results}, we present our findings, starting with the analysis of interface states in AGNR heterojunctions for different bonding configurations (\Secref{sec:int-states}).
In \Secref{sec:Nint_for_types} we systematically analyze the number of states at different junction interfaces for a wide variety of width combinations.
In \Secref{sec:effective} we derive an effective model to predict the number of interface states related to the interface geometry.
We then explore how interface states behave under geometrical deformations by applying a uniaxial-strain (\Secref{sec:geometry-deformation}), followed by an examination of the magnetic ground state (\Secref{sec:magnetism}) and the first excited magnetic states (\Secref{sec:first_excited}) at the interface of these heterojunctions. Finally, \Secref{sec:conclusions} summarizes the key findings of the manuscript and Appendix~\ref{sec:DFT} presents first-principles DFT calculations that benchmark the TB-MFH results for the magnetic properties and strain-induced magnetic transitions at the junctions.

\section{
Computational framework}
\label{sec:methods}

The electronic properties of the AGNR heterojunctions (depicted in \Figref{fig:heterojunction-types}) are modeled using the well-established one orbital MFH Hamiltonian \cite{Hubbard1963}, which has proven effective in describing these $sp^2$ hybridized systems according to experimental results and other more accurate descriptions, such as DFT calculations \cite{FernandezRossier2007, Yazyev2010, Hancock2010, Li2019}.
The Hamiltonian of these junctions can be written as a sum of the Hamilltonian corresponding to the left and right AGNRs ($H_{\ell}$ and $H_{r}$, respectively) and the coupling between them ($V$), \ie, $H=H_\ell+H_r+V$. In terms of microscopic parameters, the Hamiltonian is expressed as:
\begin{equation}\label{eq:Hubbard}
\begin{split}
    H & = \sum_{\substack{<ij>\in \ell,\\ \sigma}}tc^{\dagger}_{i\sigma}c_{j\sigma} + U\sum_{\substack{i\in\ell,\\ \sigma}}\langle n_{i\sigma} \rangle n_{i\overline{\sigma}} \\
    & + \sum_{\substack{<ij>\in r, \\ \sigma}}tc^{\dagger}_{i\sigma}c_{j\sigma} + U\sum_{\substack{i\in r, \\\sigma}}\langle n_{i\sigma} \rangle n_{i\overline{\sigma}} \\
    & +\sum_{\substack{i\in \ell, j\in r, \\ \sigma}}t_{\mathrm{int}}\left(c^{\dagger}_{i\sigma}c_{j\sigma}+c.c.\right),
\end{split}
\end{equation}
where $U$ represents the on-site Coulomb repulsion parameter, accounting for the electron-electron interaction within the same $p_z$ orbital. The operators $c_{i\sigma}$ and $c^{\dagger}_{i\sigma}$ denote the annihilation and creation operators, respectively, of an electron at site $i$ with spin index $\sigma=\lbrace \uparrow,\downarrow\rbrace$, while $n_{i\sigma}=c^{\dagger}_{i\sigma}c_{i\sigma}$ is the corresponding number operator.
We adopt the first-nearest neighbor (1NN) TB model where the hopping parameter is $t=2.7$ eV \cite{CastroNeto2009} for atoms $i,j\in \ell, r$ separated by a first-neighbor distance of $d_{<ij>}=1.42$~\AA. Similarly, $t_{\mathrm{int}}$ is the hopping at the interface between the two AGNRs, which is also a first-neighbors hopping, \ie,$t_{\mathrm{int}}=0$ for $i\in \ell, j\in r$ separated by $d_{<ij>}> 1.42$~\AA.
This simple yet efficient description allowed us to systematically explore a wide variety of junctions of different widths and interface configurations.

The TB results (with $U=0$) presented below, are obtained for very large finite systems. In contrast, self-consistent calculations with $U>0$ were performed imposing open boundary conditions and employing the Green's function methodology to solve the Schr\"odinger equation. In this way, we can effectively isolate the different spin states at the junctions from interactions with magnetic moments that can appear at the other terminations of finite systems.
For this purpose, we use our custom Python library \textsc{hubbard} \cite{dipc_hubbard} which relies on \textsc{sisl} \cite{zerothi_sisl}. The details for this implementation, can be found in \cite{Sanz2022, dipc_hubbard}.
To calculate the magnetic states in the heterojunctions, we use a sufficiently long scattering region ($L=15$ unit cells for each AGNR) to ensure that the electrodes, placed at the free ends of each AGNR, exhibit bulk-like behavior. Numerically, the total spin moment is computed as
\begin{equation}\label{eq:Sz}
    S_z = \sum_i s_z(i),
\end{equation}
where the sum runs over all atoms in the scattering region, and the atomic spin polarization is given by
\begin{equation}\label{eq:sz_i}
	s_z(i) = \frac{1}{2}\left(\langle n_{i\uparrow} \rangle - \langle n_{i\downarrow} \rangle \right).
\end{equation}
Note that the sign of $s_z(i)$ depends on the arbitrary choice for spin orientation; thus, only relative spin alignments have physical significance.

Note that the quantitative details of our results depend on the specific choice of the model parameters. The on-site interaction $U=3$ eV follows standard values reported in the literature that reproduce reasonably well the energy scales obtained from DFT for localized edge and defect states  \cite{FernandezRossier2007, Yazyev2010, Hancock2010, Li2019}. Moreover, electronic correlations beyond mean-field may further renormalize the low-energy states \cite{Joost2019}, although they are not expected to alter the qualitative picture described here.

\section{End states of AGNRs}
\label{sec:end-states-AGNRs}
The emergence and features of ESs in finite-length AGNRs can be intuitively understood by decomposing their band structure into transverse modes \cite{Brey2006}, each of which can be effectively described by a of Su-Schrieffer-Heeger (SSH) chain \cite{PhysRevLett.42.1698}.
This representation, discussed in Ref.~\cite{GarciaFuente2023} and in \SIsecKREP~of the Supplemental Material (SM) \cite{SM}, enables a mode-by-mode analysis of the topological character and the associated emergence of ESs in AGNRs.
Such mapping is exact within a $\pi$-orbital TB model with 1NN interactions. The transverse modes correspond to a set of quantized wave vectors $k_n$, determined by the boundary conditions of an AGNR of width $W$ ($W$-AGNR), defined in \Figref{fig:heterojunction-types}a:
\begin{equation}
    k_{n} = \frac{2\pi n}{(W+1)a}, \ \ n=1,\ldots\lfloor{\frac{W+1}{2}}\rfloor.
\end{equation}

For odd $W$, this collection of allowed transverse modes includes 
$k_{(W+1)/2}=\frac{\pi}{a}$,  which gives rise to two non-dispersive bands in their band structure, and whose wavefunctions are strongly localized in the wider parts of the AGNR unit cell \cite{GarciaFuente2023}.
The rest of the transverse modes correspond to dispersive bands that can be effectively described by SSH-like chains with intra- and inter-cell hopping amplitudes given by $t_\mathrm{intra}(k_n) = 2t\cos(k_n\frac{a}{2})$ and $t_\mathrm{inter}(k_n) = t$, respectively (see \SIsecKREP~\cite{SM}).
From this representation, each AGNR band exhibits a nontrivial topological character its the corresponding $k_n$ satisfies
\begin{equation}\label{eq:t_topo}
    2t\cos(k_n\frac{a}{2})< t.
\end{equation}
This condition implies that the non-trivial bands correspond to those $k_n>\frac{2\pi}{3a}$.
For finite AGNRs with even $W$ there is only one possible zigzag termination. Each topological transverse mode contributes with one topological state per end and, therefore, the number of ESs ($M$) is equal to the number of topological $k_n$. 
For odd values of $W$, the
precise value of $M$ depends on the terminating unit cell of the ribbon.
As mentioned before, the topological invariant $|Z|$ is equal to $M$ \cite{Jiang2021, LopezSancho2021}.

It should be noted that the emergence of ESs in AGNRs depends on a minimum length requirement to ensure sufficient decoupling of ESs appearing at different ends of the ribbon~\cite{Zdetsis2023}. This required length is determined by the decay length associated with the respective $k_n$, and related to the corresponding band gap $E_g(k_n)=2t|1-2\cos(k_n\frac{a}{2})|$.

Here we investigate the two types of terminations depicted in \Figref{fig:heterojunction-types}, End-I [depicted in blue in panels (a,b)] and End-II [depicted in red in panel (b)]. The ES derived from $k_{(W+1)/2}=\frac{\pi}{a}$ only appears for the End-II termination, resulting in $M_{II}=M_I+1$ (see \SIsecKREP~\cite{SM}).
Because of this somewhat richer scenario, that gives rise to two different types of heterojunctions, we will focus below on odd values of $W$. However, results for End-I terminations
can be directly generalized to cases with even $W$. 

Two types of junctions are considered: Type-I (\Figref{fig:heterojunction-types}a), conformed by two End-I unit cells, and Type-II (\Figref{fig:heterojunction-types}b), conformed by one End-I and one End-II unit cells. As the two types of AGNR terminations exhibit different bulk–boundary correspondences, their combination is expected to influence both the emergence and nature of the interface states \cite{Cao2017, LopezSancho2021}.
Note that Type-II junctions only make sense for fully-coupled configurations.

\section{Results}
\label{sec:results}

\subsection{Interface states and bonding configuration}
\label{sec:int-states}
\begin{figure}[!t]
	\centering
	\includegraphics[width=\columnwidth]{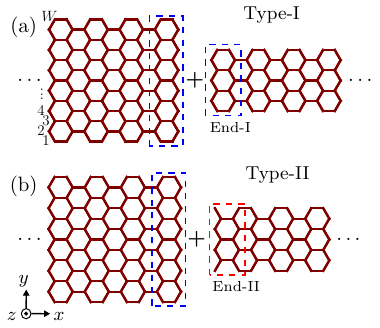}
	\caption{Atomic structure of the two possible junctions for an example case consisting of a 13-AGNR (left) and a 7-AGNR (right). (a) Type-I junction, formed by two End-I unit cells (blue dashed rectangles). The AGNR width ($W$) is defined as the number of atoms across, as indicated in panel (a).
    (b) Type-II junction, formed by one End-I (blue dashed rectangle) and one End-II (red dashed rectangle) termination cells. Inset axes in the lower part of panel (b) define the $xyz$ directions.}
	\label{fig:heterojunction-types}
\end{figure}
\begin{figure*}[t]
	\centering
	\includegraphics[width=\textwidth]{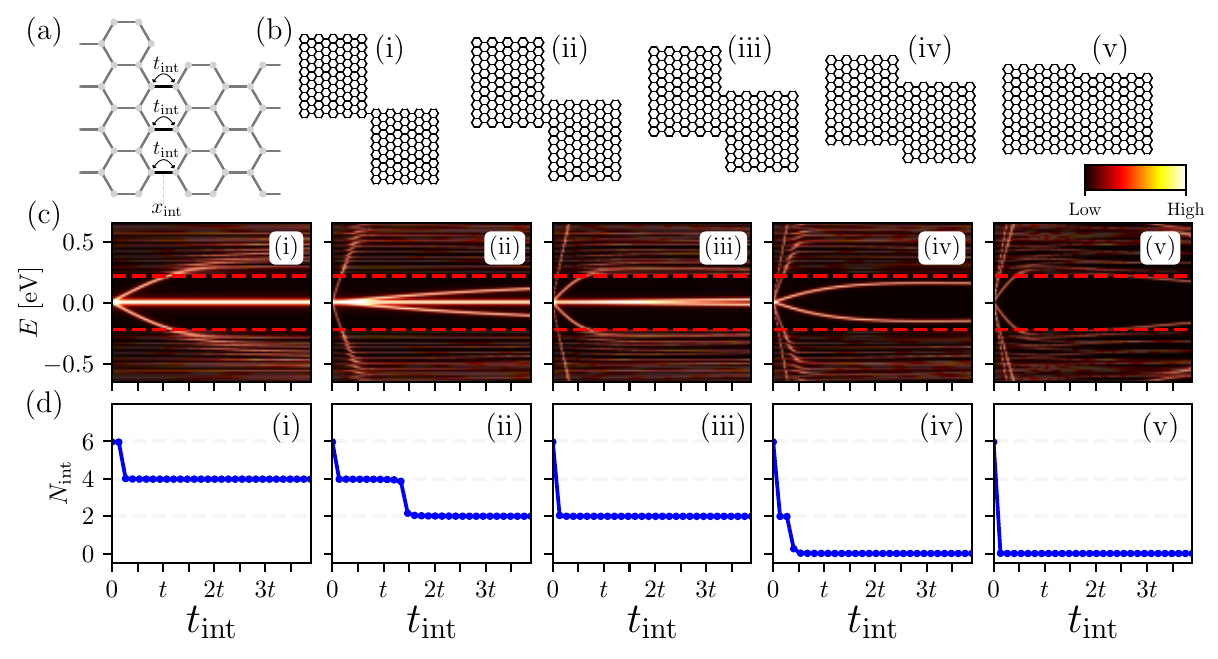}
	\caption{Interface states analysis for a finite 21-19 Type-I heterojunction with different vertical alignments. (a) Sketch of the carbon backbone structure at the junction, highlighting the interfacial hopping elements ($t_\mathrm{int}$) and the horizontal position of the interface ($x_\mathrm{int}$).
    (b) Geometries (i-v) with one to nine C-C bonds at the interface in steps of two, respectively.
    (c) LDOS map at $x_{\mathrm{int}}$ within an energy window of $\pm 0.65$ eV for each (i-v) geometry shown in panel (b) as a function of $t_{\mathrm{int}}$, with $t=2.7$ eV elsewhere.
    Red dashed lines indicate the band-gap of the system ($\pm E_g/2$). Inset colorbar indicates the low and high LDOS regions in (c). (d) Number of interface states ($N_\mathrm{int}$), as a function of $t_{\mathrm{int}}$, for each geometry (i-v) shown in panel (b), calculated as explained in \SIsecNint~\cite{SM}. These calculations were performed with $U=0$.
}
	\label{fig:interface_states}
\end{figure*}
In this section we investigate the formation of interface bound states that emerge at the junctions of AGNRs of varying interface connections, characterized by their relative vertical alignments (\ie, different number of C-C bonds at the interface).
We start by characterizing a Type-I junction formed by a 21- and a 19-AGNR (21-19) at the interface position, $x_\mathrm{int}$ (see \Figref{fig:interface_states}a). Both ribbons share the same topological classification ($Z=3$) and possess three ESs, $M_\ell=M_r =3$, where $\ell$ $(r)$ stands for the left (right) AGNR in \Figref{fig:heterojunction-types}.
We examine five different vertical alignments, where the number of C-C bonds at the interface ranges from one to nine in steps of two [see geometries (i-v) in \Figref{fig:interface_states}b].
For each geometry, we investigate the behavior of the interface states by analyzing the local density of states (LDOS) at $x_{\mathrm{int}}$, in \Figref{fig:interface_states}c.
We evaluate the LDOS as a function of the coupling between the two ribbons, $t_{\mathrm{int}}$ (see \Figref{fig:interface_states}a), within an energy window of $\pm 0.65$ eV.
In this sense, for $t_\mathrm{int}=0$, we recover the results for two finite unconnected ribbons (six states at the junction resulting from the contribution of $M_\ell$ and $M_r$), while as $t_\mathrm{int}$ increases, some of these states hybridize and move away from zero energy. 
More details for these calculations are provided in \SIsecNint~\cite{SM}.

For configurations with fewer bonds (one to three), we observe high LDOS at $E=0$ irrespective of $t_\mathrm{int}$ in \Figref{fig:interface_states}c(i-iii), indicating the presence of unhybridized zero-energy modes at the interface. In contrast, configurations with higher connectivity (more than seven bonds) display no zero-energy states, as all localized states are hybridized at the interface, as seen in \Figref{fig:interface_states}c(iv-v).

Following the dispersion of the different peaks as a function of $t_\mathrm{int}$ is also quite informative.
For instance, in panel \Figref{fig:interface_states}c(i) two states move away from $E = 0$ as $t_{\mathrm{int}}$ increases, indicating partial hybridization, while a persistent peak at $E = 0$ reflects remaining unhybridized states. 
The two dispersing peaks eventually leave the in-gap region (delimited by
the red-dashed lines), losing intensity progressively.
Note that, in absence of Coulomb interactions (\ie, $U=0$), electron-hole symmetry is perfectly preserved.
In panel \Figref{fig:interface_states}c(ii) we observe four states lying at $|E| > 0$, meaning two from each AGNR hybridize, while the remaining ones 
stay at zero energy. Two of them hybridize very strongly, leaving the in-gap region for $t_\mathrm{int}\sim t/2$, while the other two exhibit a slower linear dispersion, developing an appreciable splitting for $t_\mathrm{int}\sim 3t/2$.
In panel \Figref{fig:interface_states}c(iii) there are four states that hybridize more rapidly than in panel \Figref{fig:interface_states}c(ii), as we observe that the second set of lines abandon the in-gap energy window for $t_\mathrm{int}\sim t/2$.
The two remaining states show some degree of hybridization but their energy splitting remains negligible over the whole studied range of $t_\mathrm{int}$.
As the number of C–C bonds increases further, zero-energy LDOS features vanish due to the energy splitting of the interface states.
This is evident in panels \Figref{fig:interface_states}c(iv-v), where all six interface states are hybridized, and no zero-energy features remain. In the case of geometry (v), the two AGNRs are fully coupled with nine interfacial bonds. In this case we observe that all states have merged the continuum bands for relatively low values of $t_\mathrm{int}$.
As expected, our results reveal that the number of states that hybridize and the pace at which they move away from zero energy depend on the precise connectivity between the AGNRs.

To quantify the number of states at the interface, $N_{\mathrm{int}}$, we sum the projected density of states (PDOS) around the junction area, accounting for the decay length of the ESs, and integrate within a narrow energy window of $\pm 50$ meV around $E=0$.
The choice of this energy cutoff ensures that each state remains spectrally isolated from the bulk continuum while capturing its full spectral weight. For example, in the 21–19 heterojunction, the smallest band gap among the two ribbons is that of the 21-AGNR, with $E_g \sim 440$ meV, making the chosen window well within the gap. Further computational details are provided in \SIsecNint~\cite{SM}.

In \Figref{fig:interface_states}d(i–v) we plot $N_\mathrm{int}$ for each geometry (i-v) and coupling strength $t_{\mathrm{int}}$. Inspecting \Figref{fig:interface_states}d(i) we observe that, for the case with one single C-C bond, there are four states at the interface for $t_{\mathrm{int}}>0$.
This result is in agreement with the LDOS analysis [\Figref{fig:interface_states}c(i)] where only two states out of six are coupled, which yields four unhybridized states at the interface.
Similarly, for the case with three C–C bonds at the interface [\Figref{fig:interface_states}d(ii)], we find that $N_{\mathrm{int}} = 4$ for $t_{\mathrm{int}} < 3t/2$, while it decreases to $N_{\mathrm{int}} = 2$ for $t_{\mathrm{int}} > 3t/2$. 
In this case, the larger contact area facilitates this hybridization, but its extent depends sensitively on the value of $t_{\mathrm{int}}$.
For the geometry with five C-C bonds, $N_{\mathrm{int}}=2$ for $t_{\mathrm{int}}>0$, as seen in panel \Figref{fig:interface_states}d(iii).
These results are again consistent with the evolution of hybridized states observed in \Figref{fig:interface_states}c(ii-iii) at different coupling strengths.
In panels \Figref{fig:interface_states}d(iv-v) we observe that $N_{\mathrm{int}}$ rapidly goes to zero, in full agreement with the observed absence of LDOS at $E=0$ in panels \Figref{fig:interface_states}c(iv-v). 
A similar analysis for 25-15 junctions is presented in \SIfigNintTwentyFive~\cite{SM}, illustrating the dependence of $N_\mathrm{int}$ on the interfacial bonding in a different case.

\subsection{Fully coupled Type-I vs. Type-II interfaces}
\label{sec:Nint_for_types}
\begin{table}[h!]
    \centering
\begin{tabular}{ c||c|c|c|c|c|c|c|c|c|c|} 
\textbf{Widths} & \textbf{3} & \textbf{7} & \textbf{9} & \textbf{13} & \textbf{15} & \textbf{19} & \textbf{21} & \textbf{25} \\
\hline 
\hline
\textbf{7} & 1 \\
\cline{1-3}
\textbf{9} & 1 & 0 \\
\cline{1-4}
\textbf{13} & 2 & 1 & 1 \\
\cline{1-5}
\textbf{15} & 2 & 1 & 1 & 0 \\
\cline{1-6}
\textbf{19} & 3 & 2 & 2 & 1 & 1 \\
\cline{1-7}
\textbf{21} & 3 & 2 & 2 & 1 & 1 & 0 \\
\cline{1-8}
\textbf{25} & 4 & 3 & 3 & 2 & 2 & 1 & 1 \\
\cline{1-9}
\textbf{27} & 4 & 3 & 3 & 2 & 2 & 1 & 1 & 0 \\
\hline
\end{tabular}
    \caption{\textbf{Type-I heterojuncion}: number of interface states
    of fully-coupled Type-I
    heterojunctions ($N_ {\mathrm{int}}^{I}$)  formed by AGNRs of different widths, calculated with $U=0$.
    The rows correspond to the left (wider) AGNR while the columns correspond to the right (narrower) AGNR, as sketched in \Figref{fig:heterojunction-types}a.}
    \label{tab:zero_energy_states_heterojunctions}
\end{table}
\begin{table}[h!]
    \centering
\begin{tabular}{c||c|c|c|c|c|c|c|c|c|} 
\textbf{Widths} & \textbf{3} & \textbf{7} & \textbf{9} & \textbf{13} & \textbf{15} & \textbf{19} & \textbf{21} & \textbf{25} \\
\hline 
\hline
\textbf{7} & 0 \\
\cline{1-3}
\textbf{9} & 0 & \cellcolor{blue!25} 1 \\
\cline{1-4}
\textbf{13} & 1 & 0 & 0 \\
\cline{1-5}
\textbf{15} & 1 & 0 & 0 & \cellcolor{blue!25}1 \\
\cline{1-6}
\textbf{19} & 2 & 1 & 1 & 0 & 0 \\
\cline{1-7}
\textbf{21} & 2 & 1 & 1 & 0 & 0 & \cellcolor{blue!25}1 \\
\cline{1-8}
\textbf{25} & 3 & 2 & 2 & 1 & 1 & 0 & 0 \\
\cline{1-9}
\textbf{27} & 3 & 2 & 2 & 1 & 1 & 0 & 0 & \cellcolor{blue!25}1 \\
\hline
\end{tabular}
    \caption{\textbf{Type-II heterojuncion}: 
    number of interface states of fully-coupled Type-II heterojunctions ($N^{II}_{\mathrm{int}}$) formed by AGNRs of different widths, calculated with $U=0$. 
    The rows correspond to the left (wider) AGNR while the columns correspond to the right (narrower) AGNR, as sketched in \Figref{fig:heterojunction-types}b.}
    \label{tab:zero_energy_states_heterojunctions2}
\end{table}

In this section, we analyze how the number of interface states  $N_{\mathrm{int}}$ depends on the junction termination (Type-I vs. Type-II) and the ribbon widths, focusing on fully coupled configurations (\ie, with the maximum number of C–C bonds). We systematically computed $N_{\mathrm{int}}$ for a wide range of width combinations, and summarize the results in Tables~\ref{tab:zero_energy_states_heterojunctions} and~\ref{tab:zero_energy_states_heterojunctions2}.

Table \ref{tab:zero_energy_states_heterojunctions} summarizes $N^\mathrm{I}_\mathrm{int}$ obtained for various fully coupled Type-I junctions formed by AGNRs of different widths.
These trends are corroborated in \SIfigWidths~\cite{SM}, where we plot $N_{\mathrm{int}}$ as a function of $t_{\mathrm{int}}$ for the junctions presented in Table \ref{tab:zero_energy_states_heterojunctions}.
Similarly, Table \ref{tab:zero_energy_states_heterojunctions2} summarizes the resulting number of interface states for Type-II heterojunctions.
Interestingly, we have verified that the results in Table~\ref{tab:zero_energy_states_heterojunctions} and Table~\ref{tab:zero_energy_states_heterojunctions2} are independent on the specific geometry of the contact and hold as far as the ribbons are fully connected.

By comparing the two Tables \ref{tab:zero_energy_states_heterojunctions} and \ref{tab:zero_energy_states_heterojunctions2}, we observe that 
in most of the cases, Type-II junctions host one fewer interface state than their Type-I counterparts.
There are a few exceptions (highlighted in blue in Table~\ref{tab:zero_energy_states_heterojunctions2}) where $N_{\mathrm{int}}^I = 0$ and yet $N_{\mathrm{int}}^{II} = 1$, as seen in junctions like 9–7, 13–15, 19–21, and 25–27. 
These results can be encapsulated in the expression  $N_{\mathrm{int}}^{II} = |N_{\mathrm{int}}^{I}-1|$.

\subsection{Effective model for interface states counting}
\label{sec:effective}

To explain the results of Tables \ref{tab:zero_energy_states_heterojunctions} and \ref{tab:zero_energy_states_heterojunctions2}, we developed an effective model that yields an expression for $N_\mathrm{int}$ in terms of the AGNR widths, termination types, and bonding configuration.

As a starting point, the coupling between the two AGNRs ($V$) is treated perturbatively, since it involves only a few hopping terms between the terminal sites of the ribbons.
In fact, the bulk states lie outside the energy band gap $E_g$, and are therefore initially separated from the ESs with an energy difference 
$\geq E_g/2$ taking into account the electron-hole symmetry of our model.
In addition, the coupling strength between dispersive bulk- and end-states
scales with the length of the AGNRs along the $x$-axis ($L$) as $\propto L^{-1/2}$, due to wavefunction normalization, further justifying the use of the first-order perturbative treatment.
Therefore, in this approach, we disregard the hybridization between ESs and extended states and consider only the coupling between ESs localized at the interface to describe the low energy physics of these heterojunctions.

For these calculations, we use $L=$40 unit cell repetitions along the $x$-axis for each AGNR (\ie, a total length of $\sim 338$ \AA), to ensure the weak coupling regime.
In this treatment, the low energy spectrum of the junction is computed from the effective Hamiltonian obtained by projecting the TB Hamiltonian of the junction in the subspace of ESs of the left and right AGNRs ($\lbrace\Phi^{\mathrm{ES}}_\ell\rbrace$ and $\lbrace \Phi^{\mathrm{ES}}_r\rbrace$, respectively). Since $\langle \Phi_\ell^ \mathrm{ES} | H_{\ell,r} | \Phi_r^ \mathrm{ES}\rangle =0$, the resulting effective Hamiltonian reads
\begin{equation}
\label{eq:V12}
    H_{\mathrm{eff}} = \begin{pmatrix}
        0 & \mathcal{V} \\
        \mathcal{V}^{\dagger} & 0
    \end{pmatrix}, \ \ \ 
\mathcal{V} \equiv \langle \Phi^{\mathrm{ES}}_{\ell}|V|\Phi^{\mathrm{ES}}_r\rangle.
\end{equation}
This effective model and the full TB spectrum are compared in \SIsecEffective\ and \SIfigSVD\ \cite{SM}, showing excellent agreement in the low energy range. Additionally, in \SIsecAnalyticalEffective\ we present the explicit expressions for $\mathcal{V}$ within our 1NN-TB model and discuss the spectrum of the effective Hamiltonian (\Eqref{eq:V12}) for Type-I and Type-II junctions of AGNRs with widths up to $W$=500.

The matrix $\mathcal{V}$ captures the effective coupling between ESs. Physically, each non-zero singular value of $\mathcal{V}$  represents a bonding/antibonding pair obtained from linear combinations of ESs from each ribbon.
The rank of this matrix, thus, gives the number of hybridizing pairs of such states.
In principle, each hybridized pair removes two zero-energy modes from the spectrum. In practice, a numerical threshold to identify the non-zero singular values is imposed in order to determine $\mathrm{rank}(\mathcal{V)}$, analogously
to the energy window needed to compute $N_{\mathrm{int}}$ in \Secref{sec:int-states}. 
Since each AGNR contributes respectively with $M_{\ell,r}$ ESs to the interface, we arrive at the following expression for the number of interface states:
\begin{equation}\label{eq:N_int}
N^T_{\mathrm{int}} = M^I_\ell + M_r^T - 2\cdot\mathrm{rank}(\mathcal{V}_T).
\end{equation}
The index $T=\lbrace I,II\rbrace$ here differentiates the junction and unit-cell type. It is omitted when unambiguous or for general features for both types. As mentioned before, in this article we only consider End-I for the left AGNR.

Moreover, $\mathrm{rank}(\mathcal{V)}$ 
reaches its maximum value for the case where the number of C-C bonds at the interface is maximum. In particular, we have observed that for fully-coupled junctions $\mathrm{rank}(\mathcal{V)}=\min(M_\ell, M_r)$. Therefore, the number of interface states boils down to:
\begin{equation}\label{eq:N_int_full}
    N^{T,\mathrm{full}}_{\mathrm{int}}=|M^I_\ell-M_r^T|,
\end{equation}
as previously inferred \cite{Jiang2021}.
The validity of \Eqref{eq:N_int_full} is further justified by the additional calculations presented in \SIsecAnalyticalEffective\ in the SM~\cite{SM}.

Eqs. \ref{eq:N_int} and \ref{eq:N_int_full} are general expressions valid for both Type-I and Type-II heterojunctions, and can be used to easily understand the results found in \Secref{sec:Nint_for_types}.
Let us first consider the case where $N_{\mathrm{int}}^I \neq 0$. For Type-II junctions (which are fully coupled by definition), the right AGNR is terminated with End-II, and hence, contains one additional ES compared to its End-I counterpart, \ie, $M_r^{II} = M_r^{I} + 1$ (see \Secref{sec:end-states-AGNRs}). This also increases the 
rank of the coupling matrix by one, $\mathrm{rank}(\mathcal{V}_{II}) = \mathrm{rank}(\mathcal{V}_{I}) + 1$.
Substituting these two expressions into \Eqref{eq:N_int}, we recover the relation $N_{\mathrm{int}}^{II} = |N_{\mathrm{int}}^{I} - 1|$.
In the special case where  $N_{\mathrm{int}}^I = 0$, the rank remains the same for both junction types, since it is limited by the smaller number of ESs, here $M^I_\ell$, since $M_r^{II} = M^I_\ell + 1$. Substituting again into Eq.~\eqref{eq:N_int}, we find $N_{\mathrm{int}}^{II} = 1$, in agreement with the data in \Secref{sec:Nint_for_types}.
This behavior is verified by explicitly computing $\mathrm{rank}(\mathcal{V}_T)$ as a function of $t_{\mathrm{int}}$, as we show for two example junctions, in \SIfigRank\ \cite{SM}.

\subsection{Interface states under applied strain}
\label{sec:geometry-deformation}

\begin{figure}[h!]
	\centering
	\includegraphics[width=\columnwidth]{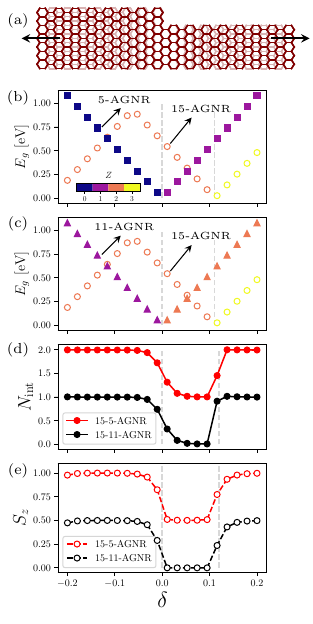}
	\caption{Topological phase space as a function of the applied strain. (a) Sketch of a strained 15-11 Type-I junction.
    (b) Band gap energy, $E_g$, as a function of $\delta$ for a 5-AGNR (filled squares) and a 15-AGNR (open circles).
    (c) $E_g$ for an 11-AGNR (filled triangles) and a 15-AGNR (open circles). The color (blue, purple, orange, yellow) identifies the topological character $Z$, as indicated by the inset colorbar in panel (b).
    (d) $N_{\mathrm{int}}$ for each $\delta$ for the 15-5 and 15-11 heterojunction, displayed in red and black lines, respectively. These calculations were performed with $U=0$. The light gray dashed lines indicate the $\delta$ values at which there are topological phase transitions.
    (e) $S_z$ calculated for $U=3$ eV for each $\delta$ for the 15-5 and 15-11 heterojunction, displayed in red and black lines, respectively.}
	\label{fig:interface_states-5}
\end{figure}

The electronic structure of AGNRs of width $W=3p+2$ (with $p$ an integer)---which appear metallic within the 1NN description \cite{Son2006b, Hancock2010}---is highly sensitive to structural deformations or environmental effects such as, \eg, interaction with a surface underneath.
More accurate theoretical models have demonstrated that these ribbons are, in fact, semiconducting \cite{Son2006b, Hancock2010}, in agreement with experimental observations \cite{Kimouche2015, Zhang2015a, Lawrence2020}.
Beyond their band gap behavior, other electronic features also depend on the atomic-scale details. For instance, the topological character of a 5-AGNR depends on the relation between the hopping parameters resulting from the relaxed geometry \cite{Lawrence2020}.
Here, we investigate how applying uniaxial strain (sketched in \Figref{fig:interface_states-5}a) can alter the properties and topological nature of the AGNRs, and therefore, the precise number of interface states at the junctions.
To understand how this geometrical deformation affects the interface states, we employ a minimal model in which strain is introduced via a dimensionless parameter $\delta$ that modifies the hopping amplitude along the direction parallel to the ribbon's axis ($t'$), such that $t'=t(1 + \delta)$. We assume that only the hopping terms strictly along the $x$-direction are affected by strain.
In this sense, $\delta>0$ corresponds to compression while $\delta<0$ corresponds to elongation.
Graphene monolayers exhibit exceptionally high mechanical resilience, with experiments demonstrating that they can sustain elastic strains of up to $\sim25\%$ \cite{Lee2008}, thus enabling the design of arbitrary strain patterns in nanoscale devices. However, we note that under compressive conditions, the ribbon may deviate from planarity, and capturing such distortions would require a more sophisticated structural model. This simplified model is not meant to describe detailed structural relaxation but rather to explore qualitative behaviors.
Furthermore, strained AGNRs may be subject to enhanced electron-phonon coupling that may induce other topological transitions \cite{Calvo2018}. 

The modified hopping leads to an adjusted condition for the emergence of topological modes, analogous to \Eqref{eq:t_topo}. This expression reads (see \SIsecKREP~\cite{SM}):
\begin{equation}\label{eq:t_topo_prime}
    2t\cos(k_n\frac{a}{2}) < t^{\prime}.
\end{equation}

In \Figref{fig:interface_states-5} we explore the topological phase diagram of periodic 5-, 11- and 15-AGNRs, as well as the interface states in 15–5 and 15–11 finite heterojunctions, as a function of the applied strain.
Panels (b) and (c) show the evolution of $E_g$ as a function of the strain parameter $\delta$ for the 5- and 11-AGNR, respectively. The color identifies the topological invariant $Z(\delta)$, calculated as the number of ESs \cite{Jiang2021, LopezSancho2021}. The vertical gray dashed lines mark the critical points where topological transitions occur.
We observe that both the 5-AGNR [filled squares in panel (b)] and the 11-AGNR [filled triangles in panel (c)] undergo a topological transition at $\delta=0$, which coincides with the band gap closure. Specifically, the 5-AGNR changes from $Z(\delta<0)=0$ to $Z(\delta>0)=1$, while the 11-AGNR evolves from $Z(\delta<0)=1$ to $Z(\delta>0)=2$, given that the 11-AGNR has one more topological $k_n$ than the 5-AGNR.
This behavior can be understood from the criterion in \Eqref{eq:t_topo_prime}: as $\delta$ decreases, fewer transverse wave vectors $k_n$ satisfy the inequality, reducing the number of topological modes. Conversely, increasing $\delta$ expands the range of $k_n$ values that meet the condition.
For comparison, we also include the band gap and topological index of a 15-AGNR (open circles) in \Figref{fig:interface_states-5}(b,c), which undergoes a transition at $\delta = 0.12$, and its classification changes from $Z = 2$ to $Z = 3$.

Panel \Figref{fig:interface_states-5}d displays the evolution of $N_{\mathrm{int}}$ for a 15-5 (red curve) and a 15-11 Type-I junction (black curve) as a function of $\delta$. The vertical dashed lines indicate the topological transitions of the individual ribbons, as discussed above.
Starting with the 15-5 heterojunction, we can see that in the region $\delta<0$ there are two states at the interface, since the 15-AGNR contributes with $M_\ell=2$ while the 5-AGNR contributes with $M_r=0$, giving $N_{\mathrm{int}}=|2-0|=2$.
In the intermediate region $0 < \delta < 0.12$, the 5-AGNR transitions to $M_r = 1$, while the 15-AGNR remains at $M_\ell = 2$, resulting in $N_{\mathrm{int}} = |2 - 1| = 1$.
For $\delta > 0.12$, both ribbons have changed their topological phase, yielding $M_\ell = 3$ and $M_r = 1$, so that $N_{\mathrm{int}} = |3 - 1| = 2$.
Similarly, in the case of the 15-11 heterojunction, in the region $\delta<0$, the 11-AGNR contributes with $M_r=1$, while the 15-AGNR contributes with $M_\ell = 2$, giving a result of $N_{\mathrm{int}}=|2-1|=1$.
In the range  $0<\delta<0.12$ both ribbons share the same topological index $Z=2$, and therefore $M_\ell=M_r=2$, yielding $N_{\mathrm{int}}=|2-2|=0$ states at the interface. For $\delta>0.12$ there are $N_{\mathrm{int}}=|3-2|=1$ states at the interface, since the 15-AGNR contributes with $M_\ell=3$.

We now turn to a key consequence of the presence of localized interface states: the emergence of localized magnetic moments when Coulomb interactions are included (\ie, $U>0$) in the MFH model [\Eqref{eq:Hubbard}].
In panel \Figref{fig:interface_states-5}e we compute the magnetic moment $S_z$, calculated with \Eqref{eq:Sz} as a function of $\delta$ for a 15-5 (red curve) and a 15-11 (black curve) heterojunction obtained with $U=3$ eV. Here we observe that, as expected, the magnetic state at the interface varies with $\delta$ following the same trend as the number of interface states. In particular, we observe that $S_z(\delta)=\frac{1}{2}N_{\mathrm{int}}(\delta)$.

Furthermore, in a real experiment the sample is typically deposited on a surface, which can in some cases screen the Coulomb repulsion interactions. For this reason we calculated $S_z(\delta)$ for $U=1,2,3$ eV in \SIfigU~in the SM \cite{SM}. From this figure we observe that the same results hold even in the case where the Coulomb interaction is significantly screened.

Given the technological relevance of controlling the spin moment at heterojunctions through strain, we carried out first‑principles DFT calculations for the 15–5 junction to validate the predictions of our simplified model. These calculations, presented in Appendix~\ref{sec:DFT} and in \SIsecStrainDFT\ of the SM~\cite{SM}, are in full agreement with our earlier results. Moreover, they provide realistic estimates of the strain at which the transition occurs, confirming that it lies well within the range that graphenic systems can sustain and that can be achieved experimentally. Specifically, for the 15–5 junction, the magnetic‑moment transition takes place under applied tensile forces between 3.5 and 6~nN, corresponding to strains of 2–4\% in the 5‑AGNR segment (and roughly four times smaller strains in the 15‑AGNR segment).

This analysis highlights how geometric deformations, such as the applied strain, provide a tunable handle to control the number of interface states, and therefore the magnetic moment at the interface, in AGNR heterojunctions.

\subsection{Magnetic ground state at the interface}
\label{sec:magnetism}
\begin{figure}[t]
    \centering
    \includegraphics[width=\columnwidth]{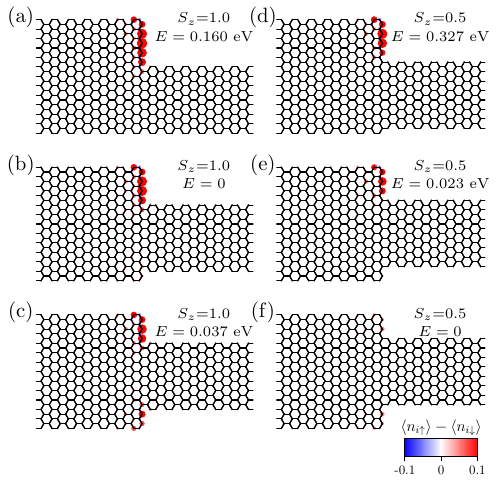}
    \caption{Spin density distribution in fully coupled 25-15 Type-I and Type-II junctions. The total energy $E$ and spin moment $S_z$ are annotated in each panel. (a-c) Type-I junctions with three different vertical alignments, respectively.
    The energy reference corresponds to the most favorable geometry shown in panel (b). (d-f) Type-II junctions with three different vertical alignments, respectively. Here we take the energy reference of panel (f) as it is the most favorable case. Size and color of the blobs at each site indicates both the magnitude and sign of the spin density, as indicated by the inset colorbar, common for all panels. These calculations were performed with open boundary conditions and $U=3$ eV.}
    \label{fig:spinpol-interface}
\end{figure}
After establishing the dependence of interface states on ribbon widths, alignments, bonding geometries, and applied strain, we performed a more exhaustive analysis on their magnetic states. In this section we study both Type-I and Type-II junctions, exploring various vertical alignments to uncover the interplay between geometry and magnetism at the interface.
The magnetic moment $S_z$ of the junctions is calculated as explained in \Secref{sec:methods} employing $U=3$ eV.

We start by computing the magnetic ground state (GS) of a junction with $N^I_\mathrm{int}\neq 0$ to unveil the magnetic properties of the  localized interface states.
For that, we choose fully-coupled junction formed of a 25- and a 15-AGNR (25-15), and consider different vertical alignments. For this junction, $N^{I}_{\mathrm{int}}=2$ and $N_{\mathrm{int}}^{II}=1$ (see Tables \ref{tab:zero_energy_states_heterojunctions}-\ref{tab:zero_energy_states_heterojunctions2}).
\Figref{fig:spinpol-interface} shows the resulting spin density distribution for 25-15 junctions, across different vertical alignments for both Type-I [panels (a–c)] and Type-II [panels (d–f)] configurations.
Both the total spin moment $S_z$ and the total energy $E$ for each junction are annotated in each panel.

Remarkably, for all alignments---including partially coupled cases (see \SIfigSpin~\cite{SM})---the magnitude of the GS spin moment remains robust and follows a clear rule:
\begin{equation}
\left|S^\mathrm{GS}_z\right| = \frac{1}{2} N^{\mathrm{full}}_{\mathrm{int}},
\end{equation}
where $N^\mathrm{full}_\mathrm{int}$ stands for the number of interface states for \emph{fully-coupled} junctions, in line with the results found in \Figref{fig:interface_states-5}e. For the 25-15 Type-I junctions, this yields $S_z=1$, while for 25-15 Type-II junctions, $S_z=0.5$. We compare our results against DFT calculations (see Appendix \ref{sec:DFT}), which confirm the validity of our approach.

In second place, we observe that the total energy of the system varies depending on the vertical alignment of the ribbons, as displayed in \Figref{fig:spinpol-interface} for both Type-I and Type-II junctions. The total energy for each configuration is given relative to the lowest-energy solution within each class of junctions.
For instance, for 25-15 Type-I junctions, geometry of panel (b) corresponds to the most energetically favorable configuration, while those in panels (a) and (c) lie 160 meV and 37 meV above, respectively. A similar trend is observed for 25-15 Type-II junctions, where the configuration in panel (f) is the lowest in energy. 
The second one is displayed in panel (e) at 23 meV, and the less favorable configuration is the geometry of panel (d) at 327 meV above the one in panel (f).

\subsection{First excited magnetic state at the interface}
\label{sec:first_excited}
\begin{figure}[t]
    \centering
    \includegraphics[width=\columnwidth]{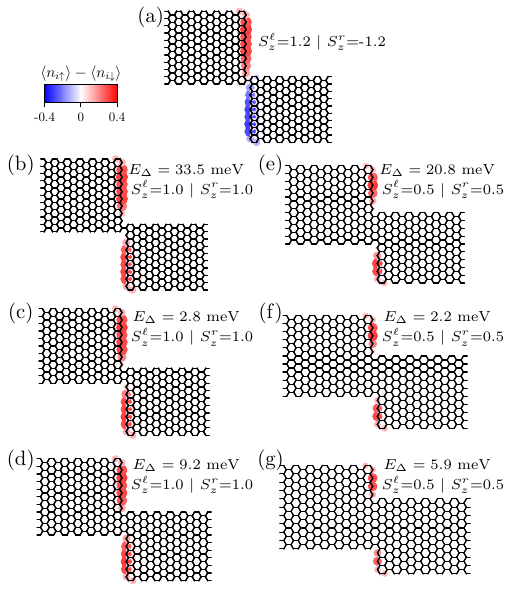}
    \caption{Spin density distribution in partially joined 21-19 Type-I junctions. (a) GS of the junction with one bond at the interface. (b-g) First excited state for different vertical alignments (one to six bonds, respectively).
    The local spins $S^\ell_z$ and $S^r_z$, and the excitation energy $E_\Delta$ for each geometry are annotated in each panel. Size and color of the blobs at each site indicates both the magnitude and sign of the spin density, as indicated by the inset colorbar, common for all panels. These calculations were performed with open boundary conditions and $U=3$ eV.}
    \label{fig:spinpol-interface-FM}
\end{figure}
We have shown in \Secref{sec:int-states} that $N_\mathrm{int}$ depends on the number of bonds at the interface. In particular, for Type-I heterojunctions composed of AGNRs with the same topological character $Z$---and thus the same number of ESs---$N_\mathrm{int}$ can still be non zero when the ribbons are only partially connected. This is due to the presence of unhybridized ESs (see \Figref{fig:interface_states}).
It is natural then, to expect emerging localized spins at the interface when the number of C–C bonds between the AGNRs is reduced.
To investigate the magnetic character of the unhybridized ESs in such junctions, we consider a 21-19 junction with a  Type-I interface, that we have already characterized in absence of Coulomb repulsion interactions in \Secref{sec:int-states}.
In \Figref{fig:spinpol-interface-FM}a we plot the GS for the case with one single C-C bond (results for the remaining 21-19 junctions can be found in \SIfigSpinAFM~\cite{SM}). 
Here we can see that local magnetic moments develop at each side of the uncoupled parts at the junction area (red and blue areas).
These local moments, labeled by $S^\ell_z$ and $S^r_z$ for the left and right ribbons, are numerically calculated by summing the atomic spin polarizations over each AGNR: $S^{\ell,r}_z = \sum_i s_z(i)$ with $i\in \ell,r$, respectively.
The color, red or blue, indicates respectively up or down spin projection. Notice that   $S_z=S^\ell_z+S^r_z$.
We observe that the magnetic GS of these junctions is $S_z = 0$, regardless the bonding configuration, in line with \Eqref{eq:Sz}. 

For the particular case of the 21-19 Type-I junction with one single bond at the interface, the local spin moments are $S^\ell_z=-S_z^r\approx1$ [\Figref{fig:spinpol-interface-FM}a], consistent with the presence of two unhybridized states in each AGNR [see \Figref{fig:interface_states}d(i)].
This kind of behavior illustrates the general validity of \Eqref{eq:Sz}.
Within each AGNR, atomic spin moments $s_z(i)$ orient in a parallel disposition, as expected~\cite{Zdetsis2020, garciafuente2024}.
This gives rise to two spin moments localized at each side of the junction with magnitude $|S_z^{\ell,r}|\approx\frac{1}{2}\left [ M_{\ell,r}-\mathrm{rank}(\mathcal{V})\right]$, corresponding to the number of unhybridized states in each AGNR.
Across the interface, these moments couple antiferromagnetically, in line with the expected alignment between opposite sublattices \cite{Lieb1989}, resulting in a total spin moment $|S_z|=|S_z^\ell-S_z^r|$=$\frac{1}{2}|M_\ell-M_r|$=$\frac{1}{2} N^\mathrm{full}_\mathrm{int}$.

Another very interesting aspect of these partially-coupled junctions is their first excited magnetic state, which depends sensitively on the number of unhybridized states.
This first excited state corresponds to a parallel alignment of $S_z^r$ and $S_z^\ell$.
Consequently, depending on the number of C-C bonds at the interface, the magnitude of the total spin of this excited state is determined by $N_\mathrm{int}$,
\begin{equation}
\label{eq:Sz_excited}
\left| S^\mathrm{exc}_z \right|= \frac{1}{2} N_{\mathrm{int}}.
\end{equation}

In \Figref{fig:spinpol-interface-FM}(b-g) we plot the spin density distribution of the first excited state for 21-19 Type-I heterojunctions of different bonding configurations. 
Here we see that, the first excited magnetic state of the junctions with one, two, and three C–C bonds at the interface [panels (b-d)] corresponds to $S_z=2$,
since $S_z^{\ell,r}$=1. This is consistent with the number of interface states $N_{\mathrm{int}}=4$ obtained from our analysis at $t_{\mathrm{int}}=t$, for the cases with one and three bonds [see \Figref{fig:interface_states}d(i–ii)].
Similarly, panels (e–g) display the excited states for junctions with four to six interfacial bonds, respectively. Here, $S^{\ell,r}_z=0.5$ for the first excited state (\ie, $S_z=1$), in agreement with our finding of $N_{\mathrm{int}}=2$ for the case with five bonds seen in \Figref{fig:interface_states}d(iii).
Junctions with higher interfacial connectivity are not shown, as we have verified that geometries with 7 or more bonds exhibit no spin-polarized states. This agrees with the absence of zero-energy states found in these cases illustrated in \Figref{fig:interface_states}c(iv–v).

We now turn our attention to the difference between the energy of the excited state ($E_{S_z}$) and that of the GS ($E_0$), $E_{\Delta}\equiv E_{S_z}-E_0$.
This excitation energy shows a highly non-monotonic dependence with respect to the geometry of the heterojunction. %
For instance, in \Figref{fig:spinpol-interface-FM}(b-d), the junctions with one, two and three bonds at the interface show 
$E_\Delta=$ 33.5, 2.8, 9.2 meV, respectively. 
Similarly, for the junctions shown in panels (e-g), the excitation energy is $E_\Delta=$ 20.8, 2.2, 5.9 meV for four, five and six C-C bonds at the interface, respectively.

Other AGNR width combinations included in the SM (\SIfigSpinFM) \cite{SM}, demonstrate that this behavior is not unique to the example discussed here.
For the junctions with one C-C bond at the interface $E_\Delta$ lies in the range $ 30-40$~meV. 
The case with two C-C bonds lies close to $\sim3$~meV, and in the case with three C-C bonds lies more or less close to $\sim 10$ meV.
The  oscillatory behavior of $E_\Delta$ continues when the number of bonds is further increased. Notice that the total magnetic moment of the excited state changes abruptly by one (following a corresponding reduction of $N_\mathrm{int}$) when the number of bonds is increased to four (up to six). Interestingly, this magnetic moment reduction is accompanied by an increase of the value of $E_\Delta$, which might be considered an unintuitive behavior.
The discussed non-trivial evolution of $E_\Delta$ applies to all the analyzed cases and may therefore represent a general feature of these junctions.

\section{Conclusions}
\label{sec:conclusions}

In summary, we have presented a detailed theoretical study of the interface states that appear in AGNR heterojunctions. By combining TB and MFH calculations, we have systematically analyzed the interplay between ribbon width, interface bonding, and geometric deformations in determining the number  of localized interface states and their magnetic properties.
In particular, we have expressed the number of interface states in terms of $M_\ell$, $M_r$, and the number of hybridized states. In a perturbative treatment, this can be calculated as the rank of the coupling matrix in the subspace of ESs of the AGNRs, $\mathcal{V}$.
For instance, for fully-coupled AGNRs, this relation boils down to $N^\mathrm{full}_\mathrm{int}$=$|M_\ell-M_r|$.
Also, the two types of junctions, Type-I and Type-II,  are related by 
$N^{II}_\mathrm{int}=|N^{I}_{\mathrm{int}} - 1|$.

We also investigated the impact of uniaxial strain applied along the junction axis using a minimal model in which the hopping parameters are altered via a dimensionless parameter. This simplified approach revealed that strain can induce topological phase transitions in AGNRs. These findings were further confirmed by first‑principles DFT calculations for the 15–5 Type‑I junction.
These transitions significantly alter the number of interface states, providing a controllable mechanism for engineering the electronic and magnetic properties of AGNR juntions. In practical settings, uniaxial strain has been experimentally applied by bending graphene deposited on flexible substrates, providing a controllable route to tune its electronic and magnetic properties \cite{Roldan2015}.
In fact, the interface magnetic moment is expected to vary with strain, in line with the strain dependence of $N_\mathrm{int}$.

In addition, our analysis of the spin states at the junctions revealed a direct correlation between $N_{\mathrm{int}}$ and the resulting spin moment at the interface, $S_z$. 
Remarkably, the total spin moment of the GS at the interface turns out to be independent on the bonding configuration, $ S^\mathrm{GS}_z$=$\frac{1}{2}|M_\ell-M_r|$=$\frac{1}{2}N^{\mathrm{full}}_{\mathrm{int}}$.
For partially-coupled ribbons, the magnetization distribution of the GS can be interpreted as the antiparallel alignment of two magnetic moments localized at each side of the junction, $S_z^r$ and $S_z^\ell$.
The lowest-lying magnetic excited state for these partially-coupled Type-I junctions corresponds to a parallel alignment of $S_z^r$ and $S_z^\ell$, with total magnetic moment $S^\mathrm{exc}_z$=$\frac{1}{2}N_{\mathrm{int}}$.
The energy of this magnetic excitation $E_\Delta$ shows a surprising non-monotonic dependence on the number of C-C bonds at the interface, which appears to be a general feature for Type-I junctions, \ie, irrespectively of the width combination, and somewhat difficult to reconcile with a Heisenberg model (\eg, a decrease of both $S_z^r$ and $S_z^\ell$ is accompanied by an increase on $E_\Delta$ when going from three to four bonds at the interface). 

The localized interface states predicted in this work are expected to produce clear experimental signatures. In scanning tunneling spectroscopy (STS), they should appear as sharp in-gap or zero-bias resonances in the dI/dV spectra \cite{Li2021}, including possible Kondo peaks when magnetic moments are present \cite{Li2019}. Moreover, their localized character also makes them ideally suited for investigation with time-resolved STM techniques \cite{Ammerman2021} and STM-based optical spectroscopies \cite{Schull2023}. In the case of partially coupled AGNRs, their low-energy spectrum may manifest as steps in the conductance, reflecting spin excitation energies \cite{Li2019}.

Our findings contribute to a deeper understanding of the electronic and magnetic properties of AGNR heterojunctions. By providing a set of predictive ``rules of thumb", this work offers a framework for designing AGNR-based devices with tunable electronic and magnetic properties. These insights pave the way for the development of nanoscale circuits and spintronic devices, leveraging the unique interplay of topology, geometry, and magnetism in GNR heterostructures.

\section*{Acknowledgements}
This work was funded by the Spanish MCIN/AEI/10.13039/501100011033 (PID2022-140854OB-C66, and JDC2022-048665-I), the University of the Basque Country (UPV/EHU) through Grant IT-1569-22.

\appendix

\section{DFT calculations}
\label{sec:DFT}

In this section, we present complementary DFT calculations of the ground-state magnetic properties of selected Type-I junctions as a function of the bonding configuration and strain. These first-principles simulations fully confirm the results obtained with the MFH model. Furthermore, the present calculations confirm that, at least in the case of the 15-5 junction, the predicted strain-induced transition takes place for strain values well below the elastic limit of this type of graphenic systems. 
\begin{figure}[hb!]
	\centering
	\includegraphics[width=0.5\columnwidth]{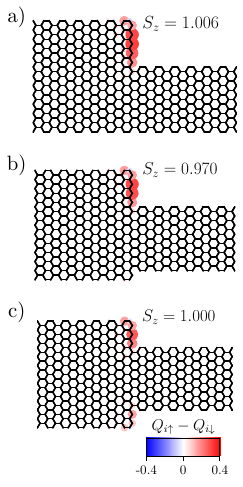}
	\caption{Atom-resolved spin polarization calculated with SIESTA for three different vertical alignments of 25-15 Type-I junctions (a-c). The magnetic moment $S_z$ at the interface is annotated in each panel. Inset colorbar is common to all panels.}
	\label{fig:spin-dft}
\end{figure}

We first consider the magnetic states of 25-15 Type-I junctions with different bonding configurations as depicted in \Figref{fig:spin-dft}.
The simulations were performed using the \textsc{Siesta} package \cite{Soler2002}, with the PBE-GGA exchange-correlation functional \cite{Perdew1996}. We used a double-$\zeta$ polarized (DZP) basis set, a mesh cutoff of 400 Ry, and an energy shift of 0.02 Ry to define the cutoff radii of the orbitals. All calculations were carried out at an electronic temperature of $T = 100$ K.
We modeled finite-length AGNR junctions comprising $L = 30$ unit cells for each AGNR (an approximate total length of 255 \AA), which ensures negligible hybridization between spin-polarized states at the interface and those at the free ends of the ribbons. All dangling bonds were passivated with hydrogen atoms.

First, the geometries were relaxed in the non-spin-polarized configuration by fixing the carbon atoms and allowing hydrogen atoms to relax until the maximum residual force was below 0.02 eV/\AA. The resulting geometries were then used in spin-polarized calculations, without further structural relaxation. Convergence for the electronic density matrix was set to $10^{-5}$ in all cases.
Numerically, the total spin moment was evaluated from the Mulliken population difference between spin-up ($Q_{\uparrow}$) and spin-down ($Q_{\downarrow}$) components,
$S_z=\frac{1}{2}\sum_i\left(Q_{i\uparrow}-Q_{i\downarrow}\right)$. 
Since in this case we are dealing with finite junctions, the sum is restricted to atomic sites $i$ with coordinates in the range $x_{\ell} \leq x_i \leq x_r$, where $x_{\ell}$ ($x_r$) denotes the $x$-coordinate of the center of the left (right) AGNR. This spatial window is chosen to account for the finite decay length of the ESs.

\Figref{fig:spin-dft} shows the DFT-computed spin polarization resolved per atom for a 25-15 junction, for three different vertical alignments [panels (a–c)], in analogy with \Figref{fig:spinpol-interface}(a-c). The spin moment localized at the interface is annotated in each case.
By comparing \Figref{fig:spin-dft}(a-c) with \Figref{fig:spinpol-interface}(a-c), and the resulting calculated spin moments, we conclude that the MFH model reproduces the more accurate DFT description.
Note that the short distance between some of the hydrogen atoms at the interface gives rise to structural relaxations and other effects that are not captured at the MFH level, precluding a meaningful comparison of the total energies. 
\begin{figure}[hb!]
	\centering
	\includegraphics[width=\columnwidth]{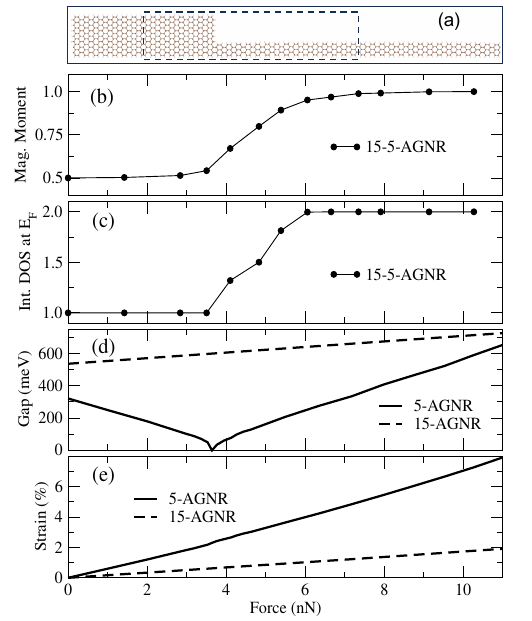}
	\caption{(a) Sketch of the unit cell used for the DFT calculation of the effect of strain in a 15-5 Type-I junction. The system is composed by a periodic arrangement of 15- and 5-AGNR segments. The highlighted region contains the carbon atoms used to define the properties of the central 15-5 junction. Panel (b) shows the evolution of the magnetic moment (S$_z$) inside such region as a function of the applied tensile force using spin-polarized calculations. Panel (c) shows the corresponding evolution of the PDOS on the carbon atoms in the central region integrated in a window of $\pm$40~meV around E$_F$ in non-spin-polarized calculations. Finally, panels (d) and (e) show, respectively, the evolution of the energy gap and the strain for the individual infinite nanoribbons as a function of the applied tension.   }
	\label{fig:strain-dft}
\end{figure}

We now turn our attention to the evolution of the magnetic moment of a 15-5 Type-I junction as a function of strain. In this case, we model the system using a superlattice formed by alternated segments containing 15 unit cells of the 15-AGNR and 30 unit cells of the 5-AGNR. The unit cell of this superlattice is depicted in \Figref{fig:strain-dft}a, has a relaxed lattice parameter of $\sim$196~\AA\ and contains 940 atoms (750 C atoms). Each unit cell contains two 15-5 junctions that, however, are sufficiently far away to assume small interactions between their associated spin moments (at least for system configurations exhibiting a non-negligible band gap). Calculations are initialized assuming an antiparallel alignment of the spin moments at the two 15-5 interfaces. Thus, the total magnetic moment in the unit cell remains always zero. However, we can define the local magnetic moment (S$_z$) associated with a given junction by integrating the spin polarization of the atoms closer to that particular interface (see the highlighted region in \Figref{fig:strain-dft}a for the case of the central 15-5 junction in the unit cell. 

Most of the technical parameters of the DFT calculations in \Figref{fig:strain-dft} are similar to those described above in the case of the 25-15 junction. In this case, however, we used a smaller electronic temperature of 30~K (identical results  were obtained lowering it further to 10~K for fixed geometries), as well as a smaller double-$\zeta$ basis (DZ) with slightly more extended orbitals (obtained with an energy shift~\cite{Soler2002} of 150~meV). This reduced DZ basis was chosen due to the large system size and the need to fully relax the structure at each applied strain value. A similar methodology—using \textsc{Siesta} together with a comparable DZ basis—has previously been shown to accurately capture the strain‑dependent evolution of the spin moment of substitutional Ni defects in graphene.~\cite{Santos2012} $\Gamma$-point  calculations were used for the large superlattice, whereas for the isolated infinite ribbons we employed a k‑point sampling of 100 points along the ribbon axis.

We first determined the equilibrium atomic structure and lattice parameter of the system shown in \Figref{fig:strain-dft}a. The lattice parameter was then progressively increased, allowing only the internal atomic coordinates to relax at each step. From the resulting stress tensor, we obtained the tensile force applied to the system. Finally, we performed spin-polarized calculations for the relaxed geometries. The results for $S_z$ at each strain value can be found in \Figref{fig:strain-dft}b. At equilibrium, the system shows $S_z=\frac{1}{2}$.w However, for an applied tension for between 3.5 and 6~nN, the system undergoes a transition to a higher magnetic moment $S_z=1$. This transition is also observed in the evolution of the number of in-gap states interface states ($N_{int}$) in \Figref{fig:strain-dft}c (where we used data from non-spin-polarized calculations for convenience): $N_{int}$ increases from 1 to 2 in the same range of applied tensions (3.5-6~nN). These calculations reproduce well the results found using the TB-MFH model [see \Figref{fig:interface_states-5}(d,e)].

The origin of these transitions becomes clear when examining the evolution of the band gaps of the infinite ribbons as a function of the applied tensile force, shown in \Figref{fig:strain-dft}d. While the band gap of the 15‑AGNR exhibits only a slight increase across the entire range of applied deformations, the behavior of the 5‑AGNR is markedly different. In this case, the gap initially decreases and closes completely at an applied tension of approximately 3.6~nN. For larger deformations, the gap reopens and increases linearly with the applied force/strain. As anticipated from the TB calculations, this closing and reopening of the band gap signals the occurrence of a topological transition. The unstrained 5‑AGNR is topological and features one ES at the End‑I termination. However, once sufficiently stretched, it becomes trivial, and no ES appears at the End‑I termination. Meanwhile, the topology of the 15‑AGNR band structure remains unchanged throughout the entire range considered in \Figref{fig:strain-dft}d, consistently contributing two ESs. Therefore, according to \Eqref{eq:N_int_full} the number of interface states at the 15-5 junction changes from $N_{int}$=1 for small strain values to $N_{int}$=0 at large deformations. Correspondingly, the magnetic moment at the interface evolves from $S_z=\frac{1}{2}$ to $S_z=1$, in full agreement with the predictions of our simplified model in Sec.IVD. More information about the spatial distribution of the interfacial states and the associated magnetic moments can be found in \SIfigStrainDFT\ in \SIsecStrainDFT\ of the SM~\cite{SM}.

Furthermore, comparing the information in \Figref{fig:strain-dft}(d,e) to that in \Figref{fig:interface_states-5}b, we can now relate the parameter $\delta$ of our model to realistic estimations of the relevant range of applied strains: the unstrained 5-AGNR corresponds to $\delta\sim+0.065$, whereas a strain of 8\% corresponds $\delta\sim-0.13$. The topological transition in the infinite 5-AGNR takes place for a strain of 2.3\%. In our periodic model of the 15-5 junction, we observe the corresponding magnetic transition at strain values in the range of $\sim$2–4\% for the 5‑AGNR segment and at values roughly four times smaller for the 15‑AGNR segment.

%



\section*{Supplementary material}

# Supplementary information: Predicting interface and spin states in armchair graphene nanoribbon junctions

Sofia Sanz[1, *] and Daniel Sánchez-Portal[1, †]

[1]*Centro de Física de Materiales (CFM) CSIC-UPV/EHU, E-20018, Donostia-San Sebastián, Spain*

**CONTENTS**

---

[*] sofia.sanzwuhl@ehu.eus

[†] daniel.sanchez@ehu.eus

## S1. TOPOLOGICAL BANDS AND END-STATE COUNTING IN AGNRS

The 1NN $\pi$-TB provides a simple way to understand the electronic properties of graphene. Using a rectangular supercell containing four atoms, as depicted in Fig. S1a, and applying periodic boundary conditions along the zigzag direction, the 2D problem gets mapped into a collection of disjoint 1D systems, as sketched in Fig. S1b. These 1D systems show two alternating hopping parameters, one of which depends explicitly on the Bloch vector along the zigzag direction $t'(k_Z)$, while the other maintains the original value $t$, yielding an SSH-like model. Cutting graphene along the zigzag edge effectively breaks bonds between sites 1 and 2, or equivalently between sites 3 and 4. In either case, according to the well-known physics of the SSH model, we can expect the appearance of an edge state whenever the condition

$$t > t'(k_Z) = 2t \cos(k_Z \frac{a}{2}) \tag{S1}$$

is fulfilled. The graphene lattice parameter $a$ is defined in Fig. S1a. Note that the SSH-chains in this formulation can be described with a unit-cell containing two effective atoms, highlighting the equivalence between atom pairs (1,2) and (3,4) in graphene.

By imposing confinement along the zigzag direction (Fig. S1a), as highlighted in blue in Fig. S2, the graphene continuum band structure becomes quantized into discrete transverse modes ($k_n$) that form the AGNR band structure [1]. These boundary conditions lead to the

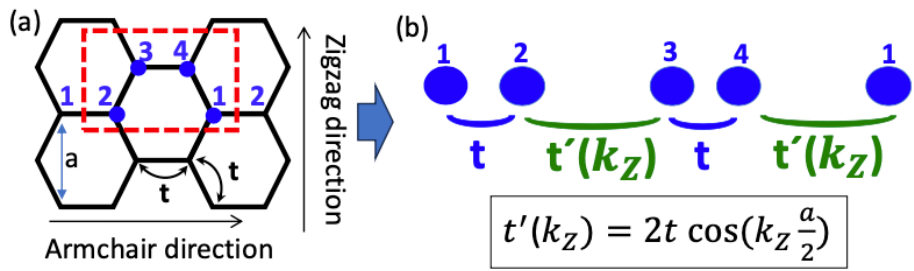

FIG. S1. (a) Rectangular graphene supercell containing four atoms with the axes oriented along the armchair (horizontal) and zigzag (vertical) directions. Red dashed lines delimitate the rectangular unit-cell. (b) Mapping of the graphene rectangular unit cell onto an SSH-like chain with two alternating hopping parameters $t'(k_z)$ and $t$.

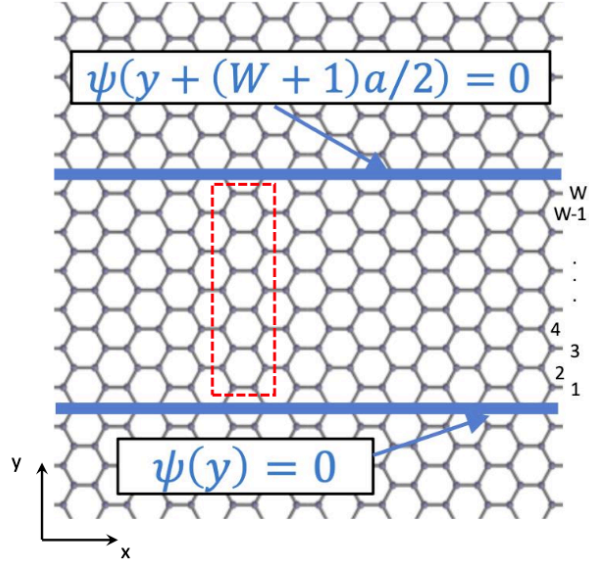

FIG. S2. Sketch of the real-space quantization conditions that must be imposed to a graphene layer to obtain the $W$-AGNR wave function $\psi(y)$. The $W$-AGNR unit cell is represented by the red rectangle.

following allowed transverse modes [2]:

$$k_n = \frac{2\pi n}{(W+1)a}, \text{ with } n = 1, 2, ..., \lfloor \frac{(W+1)}{2} \rfloor, \quad (S2)$$

with $\lfloor x \rfloor$ the floor function.

The AGNR band structure can be thus described by a set of uncoupled SSH-chains with hopping parameters given by an intra-cell hopping $t_{\text{intra}} = 2t\cos(k_n \frac{a}{2})$ and an inter-cell hopping $t_{\text{inter}} = t$. Note that this procedure reproduces exactly the ribbon's band structure within a 1NN $\pi$-TB description. Accordingly, the AGNR bands can be classified by their $k_n$, which determines the number of nodes across the ribbon and its topological character. For instance, the topologically non-trivial bands appear for those $k_n$ that fulfill Eq. (S1), i.e., $t > 2t\cos(k_n \frac{a}{2})$. The global topological index $Z$ of the AGNR is determined by the number of transverse modes that fulfill this condition.

Equation Eq. (S1) implies that an ES appears whenever the quantized transverse momentum satisfies $k_n > \frac{2\pi}{3a}$ or, alternatively, $n > \frac{W+1}{3}$. As a result, the number of ESs for a $W$-AGNR is given by:

$$M = \lfloor \frac{(W+1)}{2} \rfloor - \lfloor \frac{(W+1)}{3} \rfloor. \quad (S3)$$

This equation is in fact correct for even values of $W$. However, for odd values of $W$, two

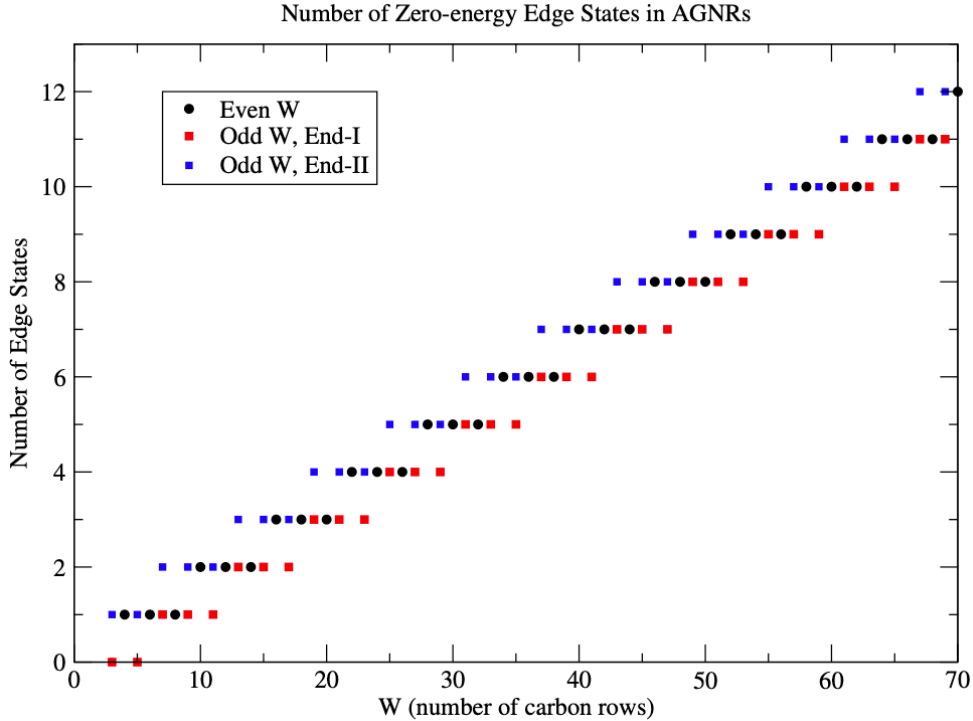

FIG. S3. Number of edge states as a function of the width $W$ of a $W$-AGNR according to Eq. (S3). Results for even $W$ are shown in black circles. Odd $W$ with End-I and End-II (see Fig. 1) are shown in red and blue squares, respectively.

distinct terminations (End-I and End-II) are possible. These two terminations are not equivalent and affect the contribution of certain modes to the ESs count. In particular, $k_{(W+1)/2} = \pi/a$, which appears only for odd $W$, is spatially localized on the wider regions of the ribbon. This mode contributes with an ES only for an End-II termination. Therefore, for odd $W$, Eq. (S3) remains valid for End-II terminations, while for End-I the number of ESs is reduced by one, yielding $M_I = M - 1$. Note that AGNRs with $W = 3p + 2$, with $p$ an integer, are metallic within the 1NN TB model. Therefore, Eq. (S3) is ambiguous for this class of AGNRs, as they are right at a topological transition point. Small changes in the relative sizes of the hopping parameters can thus change the number of ESs given in Eq. (S3) [3], as shown in Sec. IV D.

Following the same procedure used for graphene, a more detailed description of how a $W$-AGNR maps onto a set of decoupled SSH chains is sketched in Fig. S4. Here, the AGNR unit cell [panel (a)] is divided into blocks indexed by $i = \{1, 2, 3, 4\}$, each containing atoms

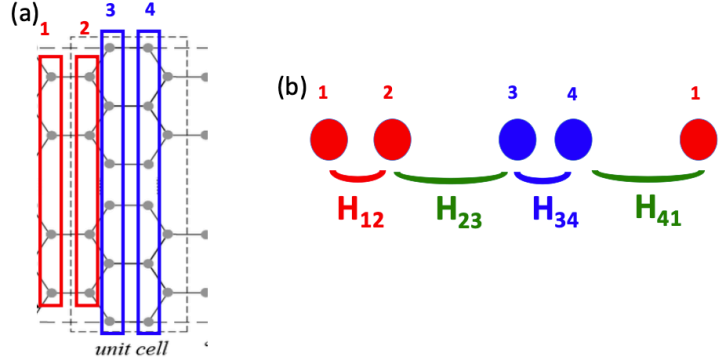

FIG. S4. (a) Mapping of the AGNR in blocks along the $x$-axis. (b) Sketch of the resulting effective 1D SSH-like chain formed by the AGNR transverse modes within each block.

located at the same $x_i$ position. The transverse modes within each $i$th block are defined as:

$$\phi_i(k_n) = \mathcal{N}_n \sum_y \sin(k_n y) p_z(i, y), \tag{S4}$$

where $p_z(i, y)$ is the $p_z$ atomic orbital located at $x_i$ and vertical position $y$. $\mathcal{N}_n = \frac{2}{\sqrt{W+1}}$ is a normalization factor for the $n$th mode, except for $k_n = \pi/a$, for which $\mathcal{N}_n = \sqrt{\frac{2}{W+1}}$. Note that $y=ma$ with $m = \{1, 2, ..., \frac{W-1}{2}\}$ for blocks 1 and 2, and $y = \frac{(2m+1)}{2}a$ with $m = \{0, 1, 2, ..., \frac{W-1}{2}\}$ for blocks 3 and 4, since blocks 1 and 2 contain $\frac{W-1}{2}$ atoms and blocks 3 and 4 contain $\frac{W+1}{2}$ atoms, respectively. As a consequence, modes $\phi_i(k_n = \frac{\pi}{a})$ for $i = \{1, 2\}$ vanish, and the SSH-like chains become a collection of disconnected (3,4) dimers giving rise to non-dispersive bands for that particular $k_n$. Since the wave function associated to this $k_n$ is localized in the dimers (3,4), this band contributes with an additional ES only for the End-II termination. Expressing the 1NN TB Hamiltonian in the basis of transverse modes, $H_{ij}(k_n) = \langle \phi_i(k_n)|\hat{H}|\phi_j(k_n)\rangle$, we find $H_{12}(k_n)=H_{34}(k_n)=$t, while $H_{23}(k_n)=H_{41}(k_n)=2t\cos(k_n\frac{a}{2})$, as expected from our previous discussion.

The uniaxial deformation along the $x$-axis considered in Sec. IV D modifies only the hopping amplitudes along bonds parallel to that direction, such that $t' = t(1 + \delta)$. In the transverse-mode basis, this translates into $H_{12}(k_n) = H_{34}(k_n) = t' = t(1 + \delta)$, while $H_{23}(k_n) = H_{41}(k_n) = 2t\cos(k_n\frac{a}{2})$. This modified hopping implies that the condition for the band to be topological [Eq. (S1)] now reads:

$$t' > 2t\cos(k_n\frac{a}{2}) \tag{S5}$$

5

## S2. CALCULATION OF $N_{\text{int}}$ AND LDOS

To evaluate the number of interface-localized states, $N_{\text{int}}$, we first compute the projected density of states per atomic orbitals $p_z(i)$,

$$\text{PDOS}(E,i) = \sum_\alpha \frac{1}{\pi} \frac{\eta_E}{(E - E_\alpha)^2 + \eta_E^2} |\phi_\alpha(i)|^2, \tag{S6}$$

where $\phi_\alpha(i)$ is the $\alpha$th eigenstate coefficient of the junction Hamiltonian at the atomic site $i$ with coordinates $(x_i, y_i)$, and $\eta_E = 0.5$ meV is the broadening parameter. We then sum over the sites $i$ around the interface and integrate over an energy window $\{-\delta E, \delta E\}$,

$$N_{\text{int}} = \int_{-\delta E}^{\delta E} \sum_i \text{PDOS}(E,i) dE, \tag{S7}$$

such that the spectral weight of each state remains well-contained within the chosen energy window, while ensuring that the states remain inside the band-gap. For the case of the 21-19 junction (Fig. 2), the band gap energy is $E_g \sim 440$ meV, for which we use $\delta E = 50$ meV, enabling an accurate count of the modes that lie well-within the gap. Furthermore, this energy scale of $\sim 100$ meV is comparable with the typical splittings induced by the spin polarization in the $U = 3$ eV calculations, allowing for a better correlation between the computed values of the spin moments $S_z$ and $N_{\text{int}}$. Note that the choice of parameters becomes of relative importance for the cases with an intermediate number of C-C bonds at the interface, since the hybridization of states evolves relatively slowly with $t_{\text{int}}$. For the extreme cases with one C-C bonds or maximum number of C-C bonds, $N_{\text{int}}$ does not show dependence on small variations of this numerical choice.

The sum of the PDOS in Eq. (S7) runs over the atomic sites $i$ such that $x_\ell \leq x_i \leq x_r$, where $x_\ell$ ($x_r$) is the $x$-coordinate of the center of the left (right) AGNR. We use these limits to account for the decay lengths of the ESs.

To calculate the LDOS map at $x_{\text{int}}$ for an energy window, we also use a spatial distribution function along $x$,

$$\text{LDOS}(E,x) = \sum_i \frac{1}{\eta_x \sqrt{2\pi}} \exp\left(\frac{-(x-x_i)^2}{2\eta_x^2}\right) \text{PDOS}(E,i), \tag{S8}$$

with a broadening of $\eta_x = 0.5$ Å.

Since $x_{\text{int}}$ lies in between atomic positions, we compute the $\text{LDOS}(E, x_{\text{int}})$ as,

$$\text{LDOS}(E, x_{\text{int}}) = \text{LDOS}(E, x_{\text{int}} - \frac{b}{2}) + \text{LDOS}(E, x_{\text{int}} + \frac{b}{2}). \tag{S9}$$

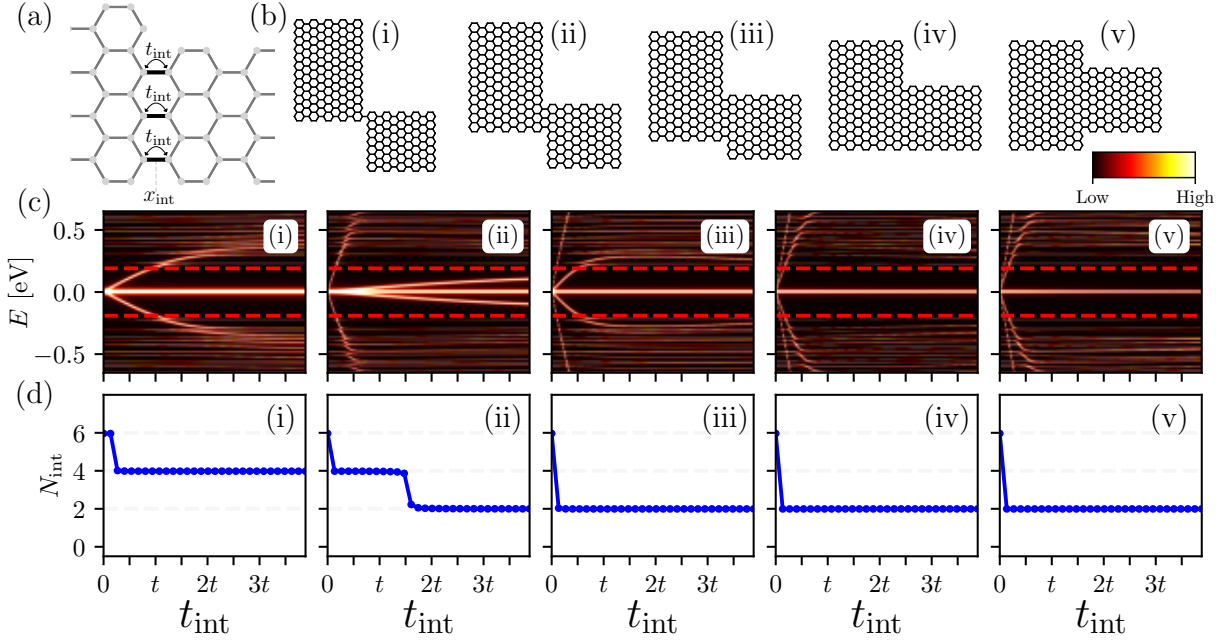

FIG. S5. Equivalent analysis of Fig. 2 for a 25-15 junction. Geometries (i-iv) have one to seven bonds at the interface in steps of two, respectively. Geometry (v) has also seven bonds but with a different vertical alignment than geometry (iv). Red dashed lines in panels (c), placed at $\pm E_g/2$, with $E_g = 381$ meV, delimitate the band gap for the 25-15 junction. These calculations were performed with $U = 0$.

Here $b = 1.42$ Å is the bond length between carbon atoms. In this way we sum the contribution from both AGNR ends at the interface. Both $N_{int}$ and $LDOS(E, x_{int})$ were evaluated for each $t_{int}$ to produce Fig. 2.

In Fig. S5 we plot an equivalent analysis of Fig. 2 for a 25-15 junction. Here we see that in this case the junction with maximum number of bonds [geometries (iv-v)] shows $N_{int} = 2$ in Fig. S5d(iv-v). We also observe that these two panels, corresponding to fully-coupled geometries but with different relative alignments, are almost indistinguishable by the eye.

## S3. SYSTEMATIC CALCULATION OF $N_{\text{int}}$ FOR MULTIPLE WIDTH COMBINATIONS

In Fig. S6 we show the calculated $N_{\text{int}}$ for fully joined Type-I junctions for the width combinations of table I, as a function of the coupling strength $t_{\text{int}}$. In other words, we show graphically the results shown in this table. As it can be seen, the number of interface states as a function of $t_{\text{int}}$ is in agreement with the results provided in table I for $t_{\text{int}} > 0$.

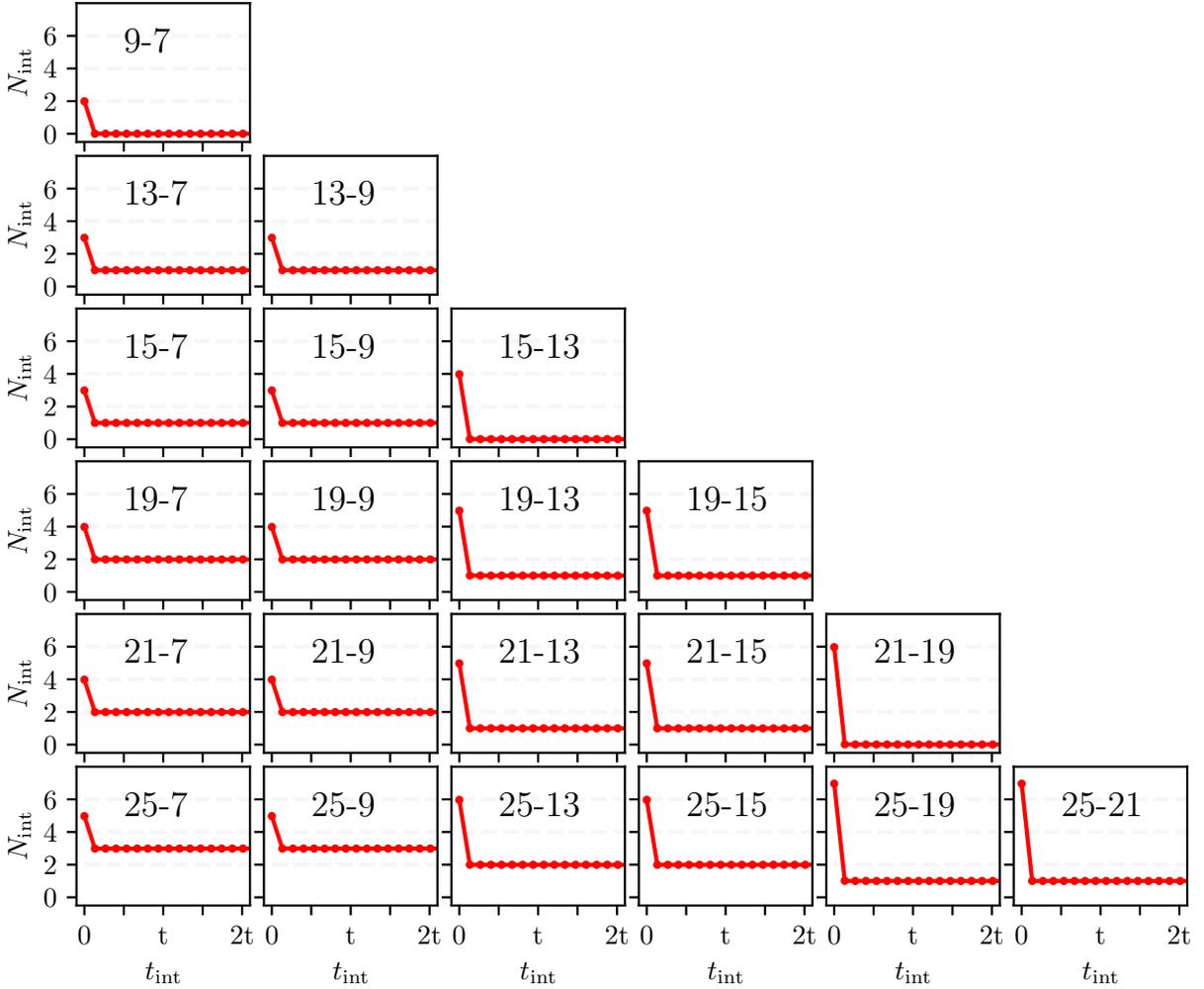

FIG. S6. Number of interface states, $N_{\text{int}}$, as a function of $t_{\text{int}}$, for different width combinations indicated in each panel. For all the heterojunctions considered here the number of C-C bonds at the interface is the maximum number allowed (fully joined AGNRs).

8

## S4. COMPARISON BETWEEN EFFECTIVE MODEL AND FULL TB HAMILTONIAN

In Fig. S7, we compare the full tight-binding (TB) spectrum of the 21–19 junction for different vertical alignments [panels (a–e)] with that of the effective low-energy Hamiltonian of Eq. (4) shown in the main text. The excellent agreement between the full TB spectrum and the spectrum of $H_{\text{eff}}$ confirms that the interaction between ESs is well described by this effective treatment, at least for the states lying within the gap (indicated by gray dashed lines in all panels), and away the strong coupling regime (*i.e.*, $t_{\text{int}} \lesssim t$). Note that, since we are dealing here with finite heterojunctions, panels (a-e) show states at $E = 0$, as the low energy spectrum contains ESs localized at the other ends of the ribbons.

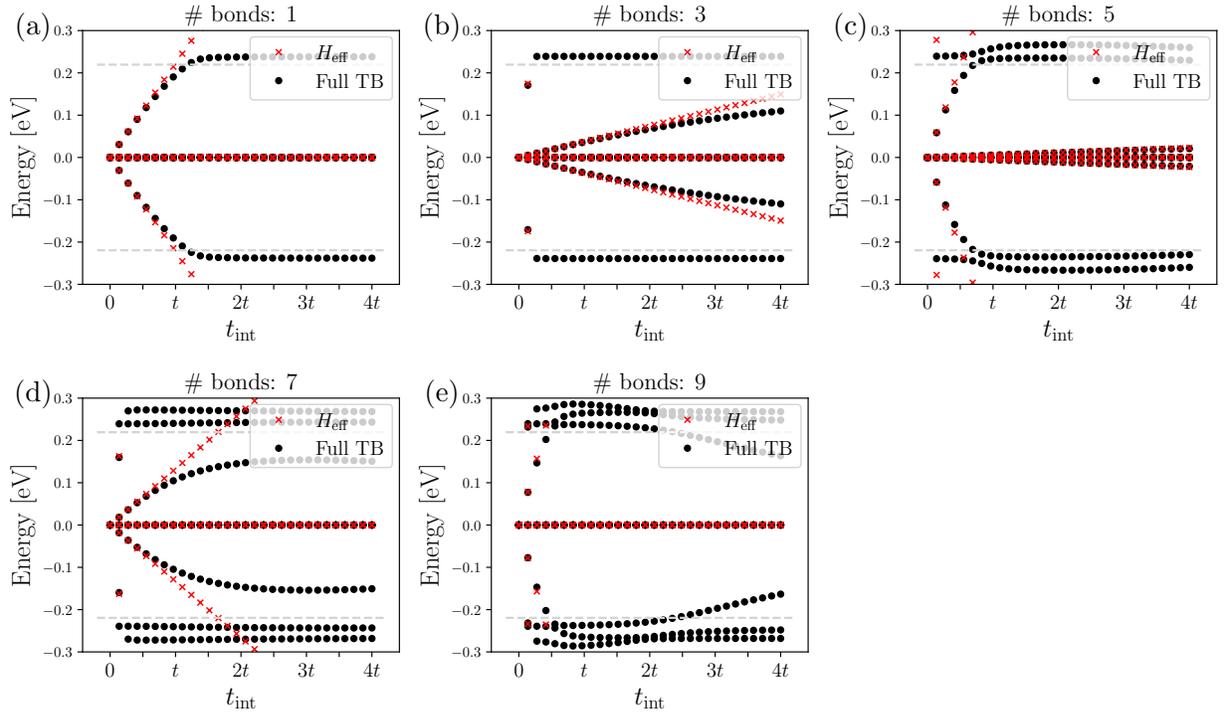

FIG. S7. Comparison of the low energy spectrum calculated from the Hamiltonian diagonalization (black dots) and with the effective Hamiltonian $H_{\text{eff}}$ (red crosses) for the 21-19 junction as a function of $t_{\text{int}}$. Panels (a-e) show results for one to nine bonds at the interface, in steps of two, respectively. These calculations were performed with $U = 0$.

## S5. RANK OF $\mathcal{V}$ FOR TWO CASE EXAMPLES

In Fig. S8 we plot the rank of $\mathcal{V}$ as a function of $t_{\text{int}}$ and interface bonding configuration for four case examples: 21-19 Type-I junction (Fig. S8a), 21-19 Type-II junction (Fig. S8b), 21-15 Type-I junction (Fig. S8c), and 21-15 Type-II junction (Fig. S8d). We consider a tolerance of $\text{tol} = \delta E = 50$ meV to calculate the rank of this matrix in order to be consistent with our calculation of $N_{\text{int}}$ (see Sec. S2).

In first place, by comparing Fig. 2 of the main text and Fig. S8a, we can clearly see that

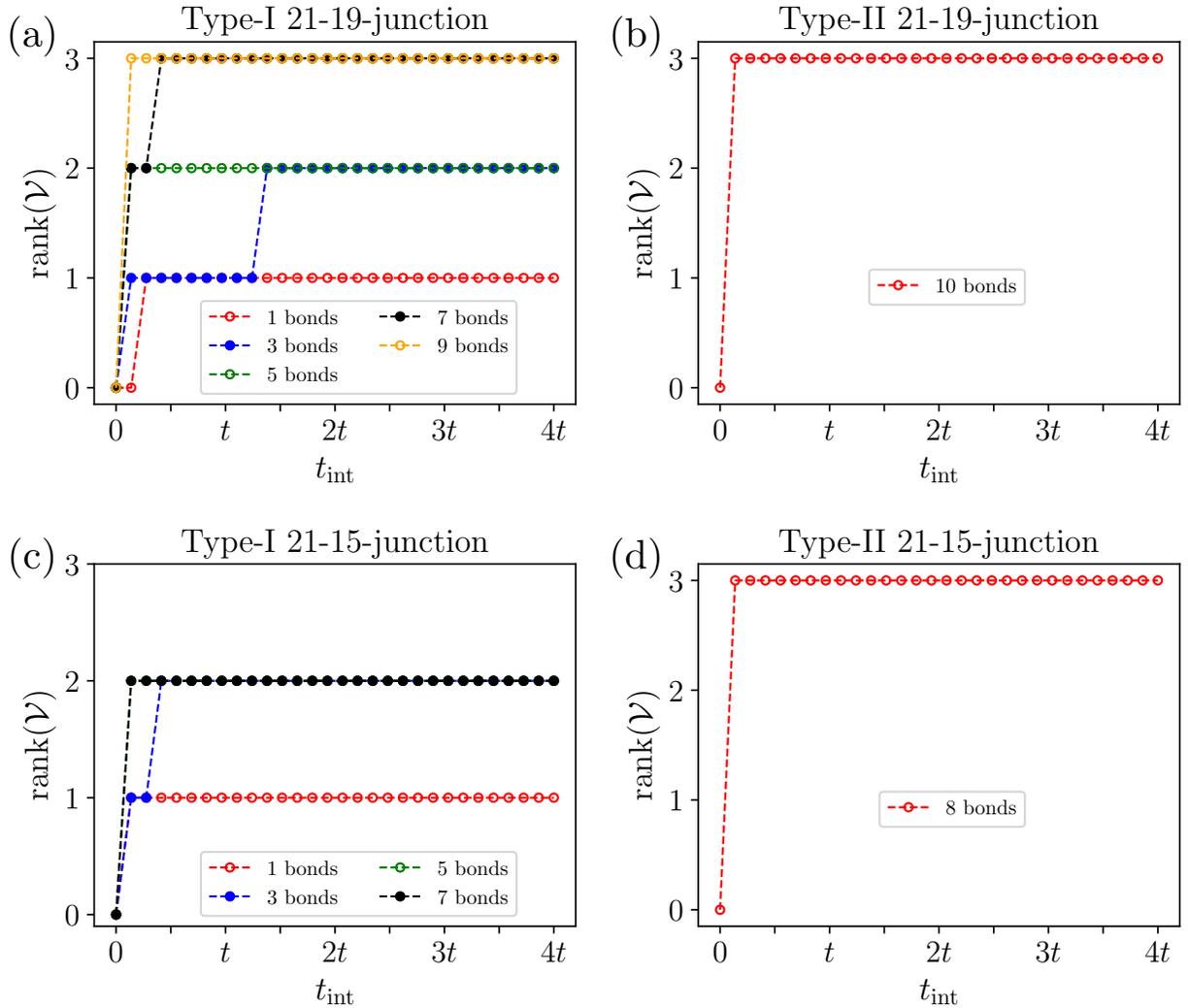

FIG. S8. Rank of the matrix $\mathcal{V}$ for a (a) 21-19 Type-I junction, (b) 21-19 Type-II junction, (c) 21-15 Type-I junction and (d) 21-15 Type-II junction, calculated for different number of C-C bonds at the junctions (as indicated in each panel's legend). These calculations were performed with $U = 0$.

the rank of the coupling matrix for each $t_{\text{int}}$ in the subspace of ESs of each AGNR explains the number of interface states for the different vertical alignments and coupling strengths.

In second place, by comparing panels Fig. S8a and Fig. S8b, we observe that $\text{rank}(\mathcal{V}_{II}) = \text{rank}(\mathcal{V}_I)$ for fully coupled 21-19 junctions (maximum number of bonds). In this case, this occurs because $M_\ell = M_r$, and therefore, the rank of this matrix is upper-limited by $\min(M_\ell, M_r)$, which in this case is the number of ESs of the End-I unit-cell (*i.e.*, $M_\ell$). However, by comparing panel Fig. S8c and Fig. S8d, we observe that $\text{rank}(\mathcal{V}_{II}) = \text{rank}(\mathcal{V}_I) + 1$ for the fully coupled 21-15 junction, as in this case $M_\ell > M_r$.

## S6. SPIN POLARIZATION FOR PARTIALLY JOINED 25-15 JUNCTIONS

Here we show the spin polarization of the magnetic GS of partially joined 25-15 Type-I junctions. In Fig. S9 we plot the spin density distribution at the junction with one to six bonds at the interface in panels (a-f), respectively. We also indicate the local spins $S_z^\ell$ and $S_z^r$, and the total energy $E$ compared to the junction of lowest energy shown in the main text (Fig. 4b). Note that the total spin $S_z = S_z^\ell + S_z^r$. As it can be seen, the total spin moment at the interface is the same ($S_z = 1$) regardless the number of bonds at the interface. Note the correlation between the calculated $N_\text{int}$ in Fig. S5 for $U = 0$ and the total spin moment $S_z$ for the different number of C-C bonds at the interface shown in Fig. S9 for $U = 3$ eV. The local spins $S_z^{\ell,r}$ calculated for each geometry shown in Fig. S9 corroborate the approximate expressions of the main text that relate these spin moments with the ESs of each AGNR, i.e., $S_z^{\ell,r} \approx \frac{1}{2}|M_{\ell,r} - \text{rank}(\mathcal{V})|$.

Focusing on the total energy we observe that, the less number of bonds at the interface leads to higher energy, as the junction gains energy by creating more bonds, as expected.

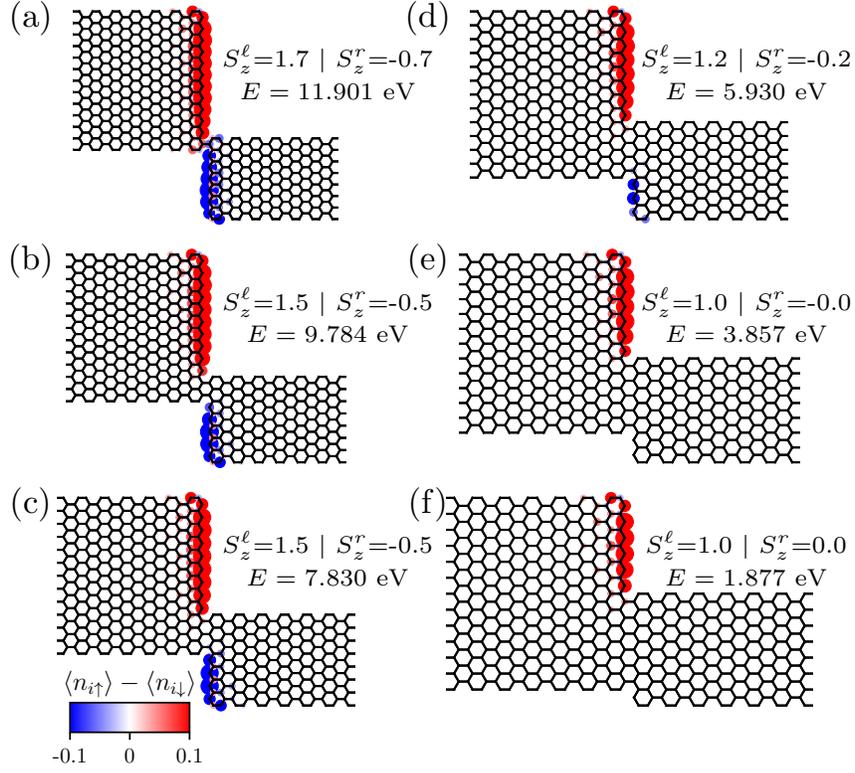

FIG. S9. Spin density distribution for the magnetic GS of the remaining partially joined 25-15 Type.I heterojunctions. The local spin moments $S_z^{\ell,r}$ are indicated in each panel. Panels (1-f) show junctions with one to six C-C bonds at the interface, respectively. Size and color of the blobs at each site show both the magnitude and sign of the spin polarization, as indicated by the inset colorbar, common for all panels. These calculations were performed with $U = 3$ eV and open boundary contidions.

13

## S7. MAGNETIC GROUND STATE OF THE 21-19 JUNCTIONS

In Fig. 5 of the main text we show the first excited state for the 21-19 partially-joined junctions. In Fig. S10 we show the magnetic GS of the same junctions in panels (a-f). We observe that, regardless the number of C-C bonds at the interface the ground state is always represented by the magnetic state with $S_z = S_z^\ell + S_z^r = 0$.

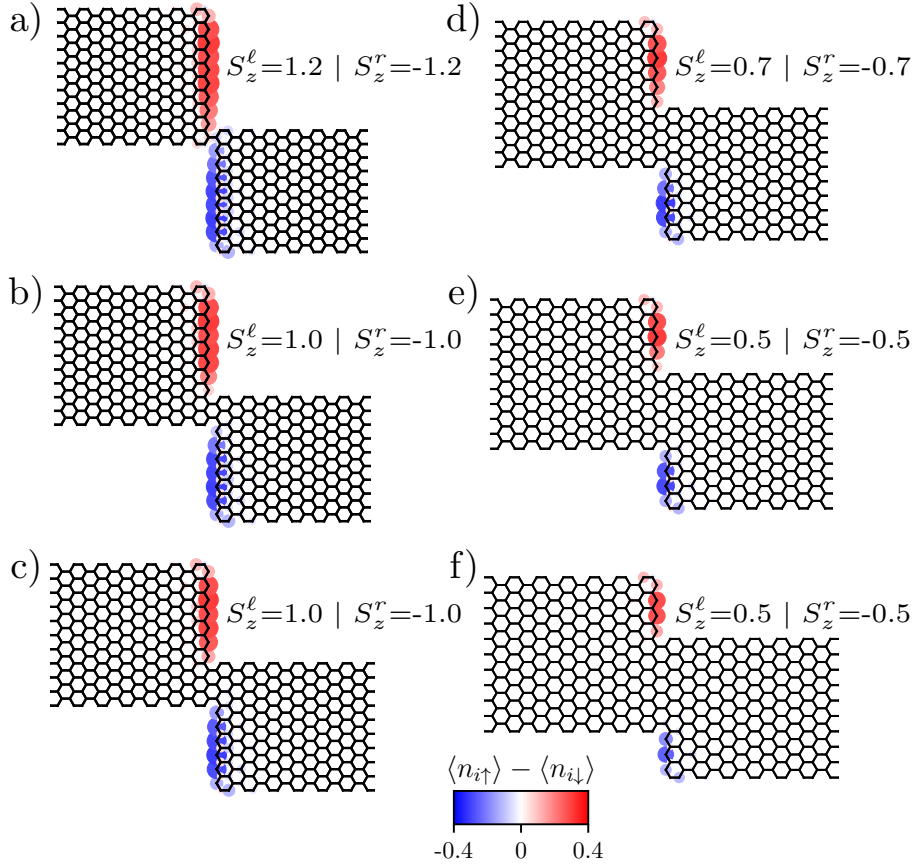

FIG. S10. Spin density distribution of the magnetic GS of the 21-19 AGNR junction for different vertical alignments (a-f). The local spin moments $S_z^{\ell,r}$ are annotated in each panel. Size and color of the blobs at each site shows both the magnitude and sign of the spin polarization, as indicated by the inset colorbar, common for all panels. These calculations were performed with $U = 3$ eV and open boundary contitions.

14

## S8. $E_\Delta$ FOR OTHER WIDTH COMBINATIONS

In the main text we presented the first excited states for 21-19 Type-I partially-coupled junctions. Here we show a similar analysis for other width combinations. As in the main text, we focus on Type-I partially-coupled junctions. We compute the first excited states and the excitation energy $E_\Delta$, for 19-15 junctions in Fig. S11, for 21-15 junctions in Fig. S12, for 25-19 junctions in Fig. S13, and in Fig. S14 for 25-21 junctions. Surprisingly, the non-monotonic growth of $E_\Delta$ with the number of bonds is not particular on the specific width combination, and persists across other junctions as well.

Again, the local spins $S_z^{\ell,r}$ calculated for each geometry shown in Figs. S11-S14 corroborate the approximate expressions of the main text $S_z^{\ell,r} \approx \frac{1}{2}|M_{\ell,r} - \text{rank}(\mathcal{V})|$.

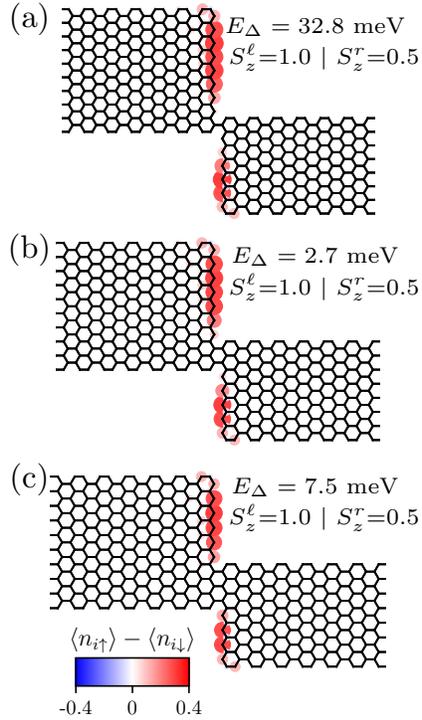

FIG. S11. Spin density distribution for the first excited state of a 19-15 heterojunction. The local spin moments $S_z^{\ell,r}$ and $E_\Delta$ are annotated in each panel. (a-c) Results for the junctions with one to three C-C bonds at the interface, respectively. Size and color of the blobs at each site shows both the magnitude and sign of the spin polarization, as indicated by the inset colorbar, common for all panels. These calculations were performed with $U = 3$ eV and open boundary contidions.

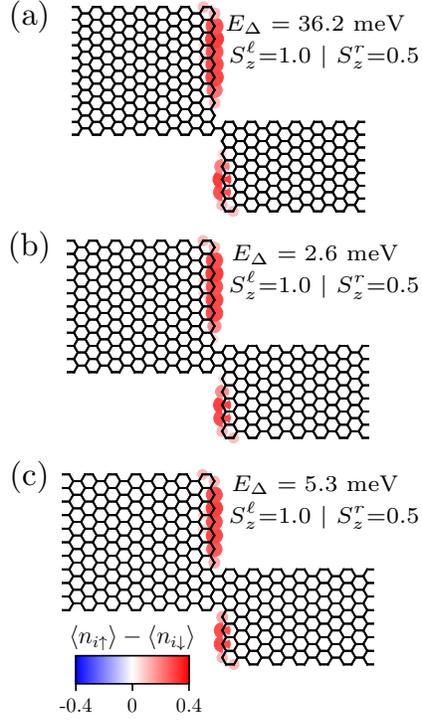

FIG. S12. Same as Fig. S11 but for a 21-15 junction.

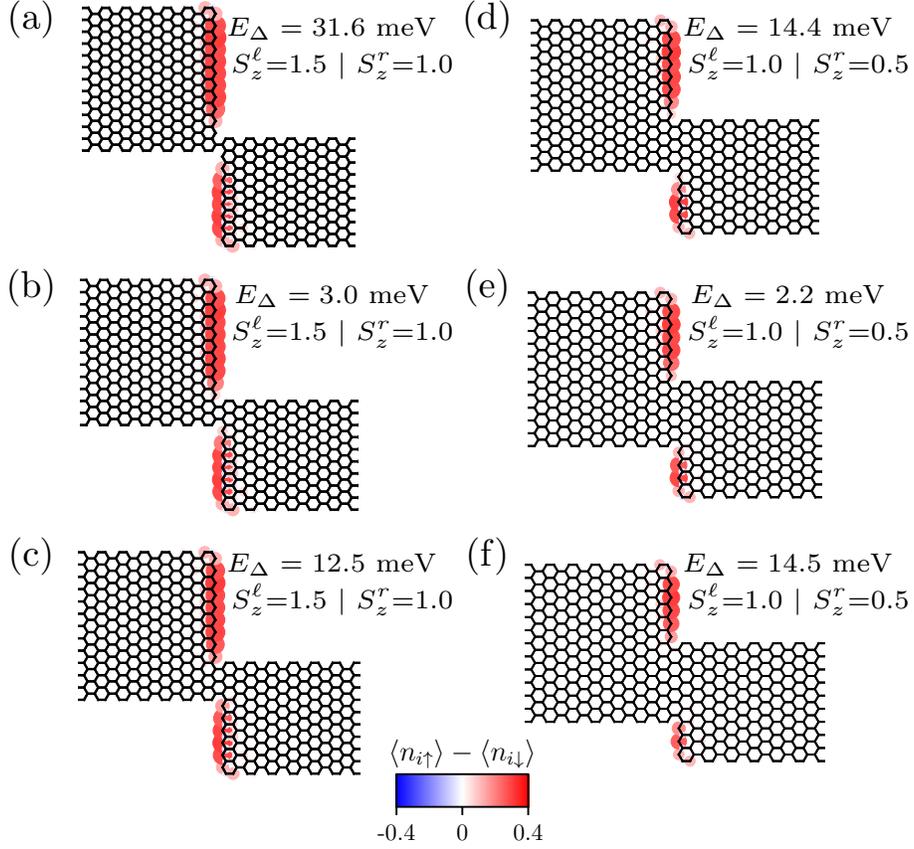

FIG. S13. Spin density distribution for the first excited state of a 25-19 heterojunction. The local spin moments $S_z^{\ell,r}$ and $E_\Delta$ are annotated in each panel. (a-f) Results for junctions with one to six C-C bonds at the interface, respectively. Size and color of the blobs at each site shows both the magnitude and sign of the spin polarization, as indicated by the inset colorbar, common for all panels.

17

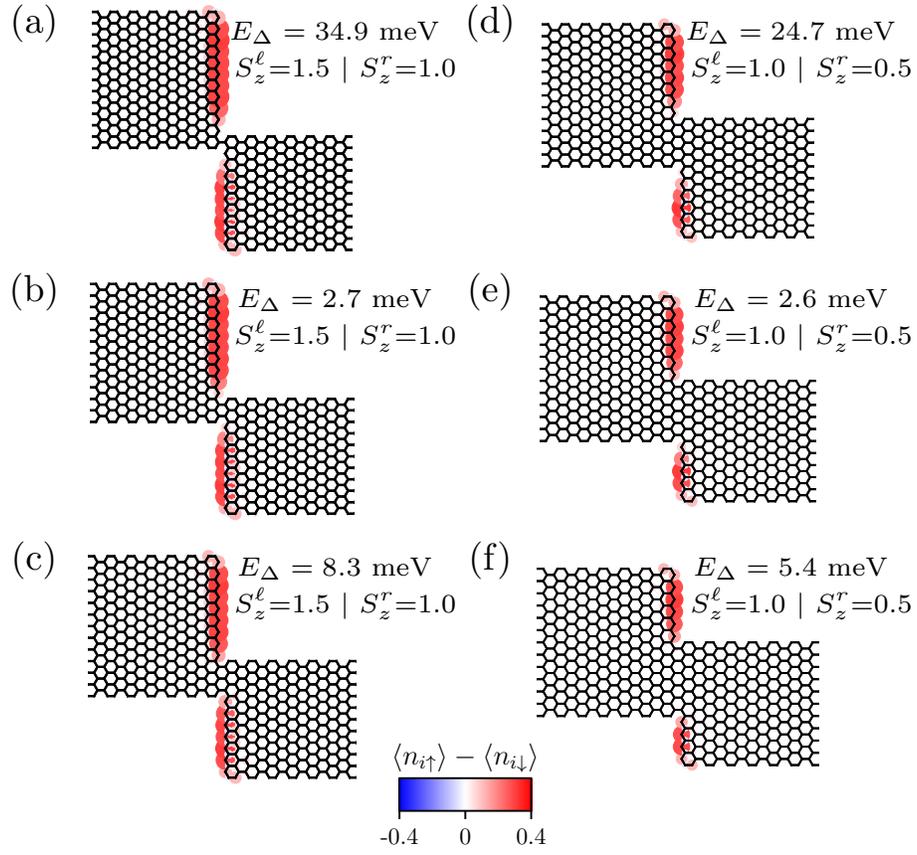

FIG. S14. Same as Fig. S13 but for a 25-21 junction.


\thispagestyle{empty}

\end{document}